\documentstyle[floats,tighten,preprint,eqsecnum,prd,aps]{revtex}

\def\diag{\mathop{\rm diag}}
\def\tr{\mathop{\rm tr}}
\begin{document}
\preprint{\parbox[t]{15em}{\raggedleft
FERMILAB-PUB-00/031-T \\ hep-lat/0002008\\[2.0em]}}
\draft

\hyphenation{author another created financial paper re-commend-ed}

\title{Application of heavy-quark effective theory
to lattice QCD:\\ I. Power Corrections}

\author{Andreas S. Kronfeld}

\address{Theoretical Physics Department,
Fermi National Accelerator Laboratory, \\
P.O. Box 500, Batavia, Illinois 60510}

\date{27 March 2000}

\maketitle 

\widetext
\begin{abstract}
Heavy-quark effective theory (HQET) is applied to lattice QCD  with
Wilson fermions at fixed lattice spacing~$a$.
This description is possible because heavy-quark symmetries are 
respected.
It is desirable because the ultraviolet cutoff~$1/a$ in current
numerical work and the heavy-quark mass~$m_Q$ are comparable.
Effects of both short distances $a$ and $1/m_Q$ are captured fully 
into coefficient functions, which multiply the operators of the usual 
HQET.
Standard tools of HQET are used to develop heavy-quark expansions of 
lattice observables and, thus, to propagate heavy-quark discretization 
errors.
Three explicit examples are given: namely, the mass, decay constant, 
and semileptonic form factors of heavy-light mesons.
\end{abstract}


\narrowtext

\section{Introduction}
\label{sec:intro}

One of the most vital parts of high-energy physics is the study of 
heavy quarks.
Several large experimental data sets of hadrons with $b$-flavored or
charmed quarks
are available now, or will be soon.
These data are valuable, because the decay properties of these hadrons 
depend on poorly known elements of the Cabibbo-Kobayashi-Maskawa (CKM) 
matrix.
A~broad range of measurements can be used to determine the CKM matrix 
with many cross checks and, thus, to test the flavor structure of
the standard model, including the origin of~$CP$ violation.

In this enterprise numerical lattice QCD plays the role of providing
hadronic matrix elements, ideally with controllable, transparent
uncertainties.
The two sources of uncertainty that attract the greatest concern are
discretization effects and the quenched approximation, in which the 
feedback of (light) quark loops on the gluons is omitted.
Both can be eliminated with ever larger computing resources.
Effects of the lattice discretization also can be studied 
theoretically.
They appear at short distances, so they can be disentangled from
long-distance physics with field theoretic methods.
This paper uses effective field theory to separate the short-distance
scales of the lattice spacing and the heavy quark mass from the
long-distance QCD scale.
It is an extension of work with El-Khadra and 
Mackenzie~\cite{El-Khadra:1997mp} on massive fermions in lattice gauge 
theory.
The effective field theory approach yields several concrete results,
which illustrate strategies for reducing cutoff effects of heavy
quarks.
Numerical calculations may then focus computer resources on
incorporating the light quark loops.

The idea of using effective field theory to study cutoff effects goes 
back to Symanzik~\cite{Symanzik:1979ph}.
For any \emph{lattice} field theory his idea was to introduce a local 
effective Lagrangian, which is the Lagrangian of the corresponding 
\emph{continuum} field theory, augmented with higher-dimension 
operators.
Coefficients of the operators depend on the underlying lattice action,
and their dimensions are balanced by powers of the lattice spacing~$a$,
which is the only short distance in Symanzik's analysis.
For small enough~$a$ one should be able to treat the higher-dimension 
terms as perturbations and express observables of the lattice theory 
as an expansion in terms of continuum observables.
For a pedagogical introduction, see Ref.~\cite{Luscher:1998pe}.

The aim of this paper is to understand numerical data generated using 
lattice actions with Wilson 
fermions~\cite{Wilson:1977nj,Sheikholeslami:1985ij} for the heavy 
quarks.%
\footnote{In this paper the term ``Wilson fermions'' encompasses
any action with Wilson's solution of the doubling
problem~\cite{Wilson:1977nj}.
These include the Sheikholeslami-Wohlert (``clover'')
action~\cite{Sheikholeslami:1985ij}, the actions of
Ref.~\cite{El-Khadra:1997mp}, and---of course---the Wilson action.}
Many papers, starting with the work of 
Gavela \emph{et al.}~\cite{Gavela:1988cf} and 
Bernard \emph{et al.}~\cite{Bernard:1988dy}, have attempted to 
calculate properties of heavy-quark systems in this way.
In practice, the bottom quark's mass in lattice units is large,
$m_ba\approx 1$, and even the charmed quark's mass is not especially
small, $m_ca\approx\case{1}{3}$.
Thus, these calculations are hardly in the asymptotic regime 
$m_Qa\to 0$ ($m_Q$ fixed) for which these actions were originally 
devised.
In particular, any expansion in small~$m_Qa$, as is usually assumed in
analyses based on Symanzik's work, fails.
This does not imply that heavy-quark cutoff effects in these 
calculations are large, but it does mean that a different analysis is 
needed.

The heavy quark masses are larger than $\Lambda_{\text{QCD}}$, so
they introduce additional short-distance scales.
One is free to seek an effective theory that lumps the effects of
all short distances---the lattice spacing and the heavy quarks'
Compton wavelengths---into coefficients.
This effective theory does not have to be continuum QCD.
The crucial observation~\cite{El-Khadra:1997mp} is that lattice
actions with Wilson fermions satisfy the same heavy-quark
symmetries~\cite{Isgur:1989vq} as continuum QCD.
For heavy-light systems, therefore, a version of the heavy-quark
effective theory (HQET) is appropriate.
Similarly, for quarkonia the same Lagrangian applies, but with the
power-counting of non-relativistic QCD (NRQCD).
The operators in such a description of lattice gauge theory are the 
same as in the usual 
NRQCD~\cite{Caswell:1986ui,Lepage:1987gg,Lepage:1992tx} or 
HQET~\cite{Eichten:1990zv,Grinstein:1990mj,Georgi:1990um,Eichten:1990vp} 
descriptions of continuum QCD, but the coefficients differ because the 
lattice modifies the dynamics at short distances.

This paper focuses on hadrons with one heavy quark and, consequently,
on~HQET.
It uses tools of the usual HQET to derive formulae of the form
\begin{equation}
	B_{\text{lat}} = z_1(m_Qa) B_\infty(\Lambda_{\text{QCD}}a) +
		\frac{1}{m_2(m_Qa)} B'_\infty(\Lambda_{\text{QCD}}a) + 
		\cdots,
	\label{eq:schematic}
\end{equation}
where~$B_{\text{lat}}$ is a physical observable calculated in lattice 
gauge theory.
The quantities~$z_1(m_Qa)$ and~$1/m_2(m_Qa)$ are short-distance
coefficients of mass dimension~0 and~$-1$, respectively.%
\footnote{Here $m_Q$ the heavy quark mass in some scheme, for example
the bare mass.  When there is more than one heavy quark in the problem,
Eq.~(\ref{eq:schematic}) is schematic, and the coefficients depend
on all heavy quark masses.}
They do not depend on the light degrees of freedom.
The quantities~$B_\infty$ and~$B'_\infty$ describe the long-distance
physics.
They are matrix elements in the infinite-mass limit and do not
depend on~$m_Q$.
Thus, the heavy-quark mass is entirely isolated into the coefficients.

The logic to derive formulae like Eq.~(\ref{eq:schematic}) parallels
that of the standard~HQET.
In both cases the deviations from the infinite-mass limit are 
expressed as a series of small corrections.
Each term consists of a short-distance coefficient multiplying a 
long-distance matrix element of the infinite-mass limit.
From this structure a simple picture of cutoff effects emerges.
The heavy-quark cutoff effects lie in the difference between the 
short-distance coefficient functions and their values in 
continuum QCD.
On the other hand, matrix elements of the infinite-mass limit, such
as~$B_\infty$ and~$B'_\infty$, suffer from discretization effects
only of the light degrees of freedom.

This paper is organized as follows:
Sec.~\ref{sec:hqet} clarifies the non-relativistic interpretation 
of Wilson fermions introduced in Ref.~\cite{El-Khadra:1997mp}, by 
giving more direct, though also more abstract, reasoning to relate 
lattice gauge theory to~HQET.
Section~\ref{sec:formalism} establishes some general notation and 
introduces the HQET Lagrangian.
As in Sec.~\ref{sec:hqet} the emphasis is on symmetries.
The leading, heavy-quark symmetric, effective Lagrangian is shown to
be the same for lattice gauge theory as for continuum QCD.
This static Lagrangian is the foundation of the heavy-quark expansion,
so some of its properties are recalled in Sec.~\ref{sec:properties}.
The next task is to propagate deviations from the static limit to
observables, so Sec.~\ref{sec:ptLI} develops a suitable form of
perturbation theory.
Applications of the formalism are in
Secs.~\ref{sec:masses}--\ref{sec:fB}.
Section~\ref{sec:masses} works out the heavy-quark expansion for
hadron masses to second order.
Semileptonic form factors, at the so-called zero-recoil point,
are addressed in Sec.~\ref{sec:semi}.
As with continuum QCD, the first order vanishes.
The technical details of the second order are considerable and
appear in Appendix~\ref{app:traces}, correcting some minor errors
in the literature.
Section~\ref{sec:fB} derives the first-order expansion for decay
constants.
In all three cases the analysis follows work on the usual HQET,
but keeping careful track of the HQET coefficients.
Some implications of these concrete results are discussed in
Sec.~\ref{sec:conclusions}.

\section{HQET for Lattice QCD}
\label{sec:hqet}

In this section lattice gauge theory, with a certain class of actions 
for Wilson fermions, is related to HQET.
A~derivation starting from the path integral of lattice QCD and making
field redefinitions has been given in Ref.~\cite{El-Khadra:1997mp}. 
That procedure is analogous to derivations
of HQET from the path integral of continuum
QCD~\cite{Korner:1991kf,Balk:1994ev,Mannel:1992mc},
and it yields the coefficients at the tree level.
Reference~\cite{El-Khadra:1997mp} used heavy-quark symmetry only
to show that the approach to the infinite-mass limit is smooth and
stable in the presence of radiative corrections.
Here the argument is reversed: owing to heavy-quark symmetry, there 
must be a version of HQET whenever momentum transfers are much 
smaller than~$m_Q$.
Whether $m_Qa\ll 1$, $m_Qa\gg 1$, or $m_Qa\sim 1$, the reasoning is 
the same.
The concepts are spelled out in this section, and the mathematical 
formalism is developed in Sec.~\ref{sec:formalism}.

To show that HQET is applicable, it is enough to show that the 
underlying theory has a set of states possessing the (approximate) 
Isgur-Wise symmetries~\cite{Isgur:1989vq}, and that no undesired 
states appear in its spectrum.
Let us consider the spectrum first, because lattice field theories 
sometimes contain spurious states.
The best-known are the doubler states.
The Wilson action employs projectors $\case{1}{2}(1\pm\gamma_4)$ to 
eliminate them completely~\cite{Wilson:1977nj}: for any three-momentum 
there is only one pole in the propagator.
Extra states also could arise if the action were to include hops over 
two or more time slices.
Without fine tuning, the propagator then would have extra poles with 
energies near~$a^{-1}$.
If $m_Qa\sim 1$, as in most numerical calculations of heavy quarks, 
the effective theory would have to describe the spurious states along 
with the desired ones.
This potential obstacle is easily circumvented~\cite{El-Khadra:1997mp}, 
however, by choosing an action with only single hops in the time 
direction.

The symmetries are revealed by writing the lattice action in the form
\begin{equation}
	S = \sum_x\bar{\psi}_x\psi_x 
	  - \kappa \sum_{x,y}\bar{\psi}_x M_{xy}\psi_y,
	\label{eq:S kappa}
\end{equation}
where $x$ and $y$ run over all lattice sites and~$M_{xy}$ has support
only for~$y$ near~$x$.
To maintain gauge invariance $M_{xy}$ includes parallel transport 
along some path from~$x$ to~$y$.
The hopping parameter~$\kappa$ controls the fermion's motion through 
the lattice, and small~$\kappa$ corresponds to large~$m_Qa$.
For $\kappa\to 0$ the propagator reduces to that of the static
theory~\cite{Eichten:1990zv}, as long as there are only single hops
in the time direction.
For small, non-zero~$\kappa$ there are spin-dependent and, with more
than one $\psi$ field, flavor-dependent corrections.
In this way, the approximate heavy-quark symmetries emerge naturally
for large~$m_Qa$.
Furthermore, if~$M_{xy}$ is constructed not to have unwanted poles 
in the quark propagator, as outlined above, on-shell Green functions 
depend smoothly on~$m_Qa$, for all 
$m_Qa>0$~\cite{El-Khadra:1997mp,Mertens:1998wx}.
Then, the Isgur-Wise symmetries persist as $a$ is reduced, and one
can always use a version of heavy-quark effective theory to describe
processes with momenta small compared to~$m_Q$.

Because the degrees of freedom and the symmetries are the same as in 
the continuum, the operators of this HQET are the \emph{same} as those 
of the usual HQET describing continuum QCD.
To define the operators the main issue is to regulate divergences.
There is no need choose the same ultraviolet regulator for the 
effective theory as for the underlying theory.
One is free to regulate the ultraviolet with, say, dimensional 
regularization and either a physical or a minimal renormalization 
scheme.
On the other hand, because the effective and underlying theories are 
supposed to describe the same long-distance physics, the same infrared 
regulator, when needed, should be chosen.

Because the details of the short-distance dynamics are those of the 
lattice theory, the coefficient functions of HQET must be modified.
The lattice breaks some rotational and translational symmetries, so at 
non-zero~$a$ coefficients of corresponding operators need not vanish, 
as they would in the usual HQET.
As~$a$ varies the short-distance properties change, and so the 
coefficients must change to compensate.
Eventually, when $a\to 0$, lattice QCD becomes (indeed, defines) 
continuum QCD, so the coefficients of the modified HQET smoothly turn 
into those of the usual HQET.
The explicit form of the coefficients is not needed in this paper, but 
one should note that they can be calculated by computing observables 
in lattice QCD and in the modified HQET and matching.
With Feynman diagrams, for example, one would expand lattice amplitudes
around the static limit in small momentum transfers, keeping the full
dependence on~$m_Qa$.

In Eq.~(\ref{eq:S kappa}) the hopping matrix~$M_{xy}$ is not specified
in detail.
In general it contains many free couplings, which are irrelevant in
the sense of the renormalization group.
In the usual improvement program~\cite {Symanzik:1983dc} they are 
chosen to accelerate the approach to the continuum limit.
In the HQET analysis advocated here, a similar principle holds.
The irrelevant couplings of the lattice action alter the 
short-distance coefficients of the modified HQET.
Thus, they can be adjusted so that the HQET expansion of lattice QCD 
systematically reproduces more and more of the HQET expansion of 
continuum~QCD.

For a generic lattice action, the heavy-quark symmetries hold only
in the rest frame.
On a superficial glance this is a drawback, because much of the power
of HQET comes from boosting heavy-light hadrons to arbitrary frames.
On a second glance, it may be a blessing in disguise.
By combining heavy-quark symmetry, Lorentz covariance, and
reparametrization invariance~\cite{Luke:1992cs}, it may be possible to
develop a non-perturbative improvement program.

\section{Notation and Formalism}
\label{sec:formalism}

This section reviews the main ingredients of HQET in a notation
well-suited to Euclidean space-time.
The details are slanted to Euclidean space-time because the aim of
the paper is to understand the output of Monte Carlo calculations of
lattice QCD.
All results, however, are for matrix elements defined at a fixed
(Euclidean) time, so they apply equally well to the Minkowski theory.
Indeed, with the conventions introduced here, the formulae in this paper 
hold for both kinds of time, unless specifically noted.

The Euclidean action can be written $S=-\int\!d^4x\,{\cal L}$, where
${\cal L}$ is the Lagrangian, and the weight factor in the functional
integral is then $e^{-S}$.
The metric is~$\delta^{\mu\nu}$, Greek indices run from~1 to~4, and
Dirac matrices satisfy $\{\gamma^\mu,\gamma^\nu\}=2\delta^{\mu\nu}$.
A~convenient basis is given in Ref.~\cite{El-Khadra:1997mp},
in particular
\begin{equation}
	\gamma^4= \left(
		\begin{array}{cc} 1 & 0 \\ 0 & -1 \end{array} \right).
	\label{eq:gamma4}
\end{equation}
As usual, we take real (Minkowski) time to be~$t=x^0$.
Then Euclidean time $x^4=ix^0$, and the general rule relating the
fourth component to the zeroth component of a four-vector~$q$ is
\begin{equation}
	q^4= iq^0.
	\label{eq:4i0}
\end{equation}
Because the spatial components are the same, it is convenient to put
all modifications into the time component.
Therefore, this paper uses the metric $g^{\mu\nu}=\diag(-1,1,1,1)$,
where Greek indices run from~0 to~3,
so $q_0=-q^0$ and $q^2=-(q^0)^2+\bbox{q}^2$.
And the Dirac matrices $\gamma^0=-i\gamma^4$ and $\gamma^j$ differ by
a factor of~$-i$ from those of the most common Minkowski convention.

The four-volume element is defined to be
\begin{equation}
	d^4x := dx^1 dx^2 dx^3 dx^4 = i\,dx^0 dx^1 dx^2 dx^3.
	\label{eq:d4x}
\end{equation}
The factor of~$i$ in Eq.~(\ref{eq:d4x}) is the most unusual convention
introduced here, but it allows many formulae given below look the
same in both Euclidean and Minkowski space-time.
For example the weight factor of the path integral is always 
$e^{\int\!d^4x\,{\cal L}}$.

The foregoing conventions can be used in any field theory.
In HQET one introduces a velocity~$v$, with~$v^2=-1$.
Although heavy-quark symmetry of lattice gauge theory is only guaranteed
in the rest frame~$\bbox{v}=\bbox{0}$, it is convenient to keep~$v$
arbitrary.
The projectors
\begin{equation}
	P_\pm(v) = \case{1}{2}(1\mp i\kern+0.1em /\kern-0.55em v)
	\label{eq:P+-}
\end{equation}
project onto ``upper'' and ``lower'' components of spinors.
For any vector~$q$ the components orthogonal to~$v$,
\begin{equation}
	q^\alpha_\perp = q^\alpha + v^\alpha v\cdot q,
	\label{eq:perp}
\end{equation}
play a special role.
In the rest frame they are the spatial directions.
It is also convenient to introduce
\begin{equation}
	{\eta^\alpha}_\beta = {\delta^\alpha}_\beta + v^{\alpha}v_{\beta}
	\label{eq:eta}
\end{equation}
to project out orthogonal components of a tensor, \emph{e.g.},
$q^\alpha_\perp	= {\eta^\alpha}_\beta q^\beta$.

In HQET heavy quarks are represented by a heavy-quark 
field~$h_v^{(+)}$ satisfying
\begin{equation}
	h_v^{(+)} = P_+(v) h_v^{(+)}.
	\label{eq:h+}
\end{equation}
The anti-quarks are represented by $h_v^{(-)} = P_-(v) h_v^{(-)}$.
As in the usual HQET one can either consider the anti-quarks to be
decoupled~\cite{Korner:1991kf,Balk:1994ev} or integrated
out~\cite{Mannel:1992mc}.
But in this paper, having shown that the heavy-quark symmetries hold
in lattice QCD, the effective Lagrangian is developed principally on
the basis of symmetry.
The heavy-quark Lagrangian is written
\begin{equation}
	{\cal L}_{\text{HQET}} =
	{\cal L}^{(0)} +
	{\cal L}^{(1)} +
	{\cal L}^{(2)} + \cdots,
	\label{eq:L}
\end{equation}
where the leading term is 
\begin{equation}
	{\cal L}^{(0)} = \bar{h}_v^{(+)}(iv\cdot D - m_1)h_v^{(+)}.
	\label{eq:L0}
\end{equation}
A non-zero \emph{rest mass}~$m_1$ is introduced to describe the
exponential fall-off of Euclidean Green functions, $e^{-E|x_4|}$ with
energies $E\approx m_1$.
The further interactions~${\cal L}^{(s)}$ contain operators
of dimension~$4+s$.
By dimensional analysis their coefficients, of dimension~$-s$, contain 
powers of the short-distance scales~$1/m_Q$ or~$a$.

The Lagrangian~${\cal L}^{(0)}$ is the unique scalar of dimension
four satisfying the Isgur-Wise symmetries~\cite{Isgur:1989vq}.
The heavy-quark spin symmetry is manifest, but with $m_1\neq 0$
the flavor symmetry is not.
It is, however, there.
In Eq.~(\ref{eq:L0}) let the field $h_v^{(+)}$ to be a column vector 
for all the flavors of velocity~$v$, and let~$m_1$ denote a mass 
matrix.
For example, for two flavors
\begin{equation}
	m_1 = \left(
	\begin{array}{cc}
		m_{1c} &   0   \\
		  0    & m_{1b}
	\end{array} \right).
	\label{eq:mmx}
\end{equation}
Let $\theta=(m_{1c}-m_{1b})v\cdot x$ and consider the generators
\begin{equation}
	\tau^1 = \frac{i}{2}\left(
	\begin{array}{cc}
		0 &  e^{i\theta}  \\
		e^{-i\theta} & 0
	\end{array} \right),\quad
	\tau^2 = \frac{i}{2}\left(
	\begin{array}{cc}
		0 & -ie^{i\theta}  \\
		ie^{-i\theta} & 0
	\end{array} \right),\quad
	\tau^3 = \frac{i}{2}\left(
	\begin{array}{cc}
		1 & 0  \\
		0 & -1
	\end{array} \right),
	\label{eq:tau}
\end{equation}
satisfying the SU(2) algebra $[\tau^d,\tau^e]=\varepsilon^{dfe}\tau^f$.
Then the flavor symmetry is
\begin{equation}
	h_v^{(+)}       \mapsto       e^{ \tau^a\omega_a}h_v^{(+)},\quad
	\bar{h}_v^{(+)} \mapsto \bar{h}_v^{(+)}e^{-\tau^a\omega_a}.
	\label{eq:hqs}
\end{equation}
The symbol~${\cal D}^\mu=D^\mu-im_1v^\mu$, which was introduced in
Ref.~\cite{Falk:1992fm}, satisfies $[{\cal D}^\mu,\tau^d]=0$ and is,
thus, trivially covariant under the transformation~(\ref{eq:hqs}).
Therefore, flavor-symmetric operators take the form
\begin{equation}
	O_\Gamma^{\mu_1\cdots\mu_n} =
	\bar{h}_v^{(+)}\Gamma{\cal D}^{\mu_1}\cdots{\cal D}^{\mu_n} h_v^{(+)},
		\label{eq:OGamma}
\end{equation}
where $\Gamma=P_+(v)\Gamma P_+(v)$.
Spin-symmetric operators have $\Gamma=1$ (or
$\kern+0.1em /\kern-0.55em v$, since
$\kern+0.1em /\kern-0.55em vh_v^{(+)}=ih_v^{(+)}$).
The only flavor- and spin-symmetric scalar at dimension four is
$\bar{h}_v^{(+)}iv\cdot{\cal D}h_v^{(+)}$, which is~${\cal L}^{(0)}$.
Thus, the symmetries of HQET with non-zero rest masses are the same as 
without.

In the following anti-quarks are not considered further, so from now on 
the heavy quark field is written~$h_v$ instead of~$h_v^{(+)}$.

To describe deviations from the symmetry limit, one introduces the 
higher-dimension interactions~${\cal L}^{(s)}$, which are built from 
operators like~$O_\Gamma$.
These are general enough to include the gluon field strength, because 
$F^{\mu\nu}=[D^\mu,D^\nu]=[{\cal D}^\mu,{\cal D}^\nu]$.
One may omit operators that would vanish by the equations of motion 
of~${\cal L}^{(0)}$, $-iv\cdot{\cal D}h_v=0$.
Such operators make no net contribution on the HQET mass shell,
so they do not appear in on-shell matching calculations.
At dimension five there can be two ${\cal D}$s, so
\begin{equation}
	{\cal L}^{(1)} =
		\frac{{\cal O}_2}{2m_2} + \frac{{\cal O}_B}{2m_B},
	\label{eq:L1}
\end{equation}
where
\begin{eqnarray}
	{\cal O}_2 & = &
		\bar{h}_vD_\perp^2 h_v, \label{eq:O2} \\
	{\cal O}_B & = & 
		\bar{h}_v s_{\alpha\beta}B^{\alpha\beta}h_v,
		\label{eq:OB}
\end{eqnarray}
with $s_{\alpha\beta}=-i\sigma_{\alpha\beta}/2$
and  $B^{\alpha\beta}=\eta^\alpha_\mu\eta^\beta_\nu F^{\mu\nu}$.
In the rest frame, ${\cal O}_2$ gives the kinetic energy and
${\cal O}_B$ the chromomagnetic interaction.
At dimension six, with three~${\cal D}$s, 
\begin{equation}
	{\cal L}^{(2)} =
		\frac{{\cal O}_D}{8m_D^2} + \frac{{\cal O}_E}{8m_E^2},
	\label{eq:L2}
\end{equation}
where
\begin{eqnarray}
	{\cal O}_D & = &
		\bar{h}_v[D^\alpha_\perp,iE_\alpha] h_v,
		\label{eq:OD} \\
	{\cal O}_E & = & -
		\bar{h}_vi\sigma_{\alpha\beta}\{D^\alpha_\perp,iE^\beta\}h_v,
		\label{eq:OE}
\end{eqnarray}
with $E^\beta=-v_\alpha F^{\alpha\beta}$.%
\footnote{The chromoelectric field of Ref.~\cite{El-Khadra:1997mp} is 
related (in the rest frame) to the one here by
$\bbox{E}_{\!\cite{El-Khadra:1997mp}}=i \bbox{E}$.}
In the rest frame, ${\cal O}_D$ gives the Darwin term and
${\cal O}_E$ the spin-orbit interaction.
The complete list of dimension-six interactions includes four-quark
operators, such as $\bar{q}\gamma^\mu q\bar{h}_vv_\mu h_v$, but their
coefficients all vanish at the tree level.

Distinct inverse masses~$1/m_2$, $1/m_B$, $1/m_D^2$, and $1/m_E^2$ are
introduced as a notation for the coefficients of the modified HQET.
One could have equally well written $z_B/m_2$ instead of~$1/m_B$,
and so on, but to trace the effects of the higher-dimension operators
on physical observables the notation of inverse masses is adequate.
The numerical factors and powers of the inverse masses have been chosen
so that all masses become the same in the tree-level continuum limit.
At non-zero lattice spacing and in the presence of radiative
corrections, this is no longer guaranteed.

Concrete expressions for the coefficients lie beyond the scope of this 
paper.
They depend on couplings of the lattice action, the
velocity~$\bbox{v}$, and the HQET renormalization scheme.
Ideally one would like to devise a non-perturbative scheme for 
computing the coefficients, but so far they have been studied only in 
perturbation theory.
For the lattice actions in common use, expressions are available at 
the tree level for~$m_1$, $1/m_2$, and 
$1/m_B$~\cite{El-Khadra:1997mp}, and at the one-loop level for~$m_1$ 
and $1/m_2$~\cite{Mertens:1998wx}.

Through dimension six the effective heavy-quark Lagrangian is 
rotationally invariant.
Starting with dimension seven, this is no longer the case.
For example, consider the term
\begin{equation}
	{\cal L}^{(3)} = \cdots +
		a^3w_4\sum_{i=1}^3 \bar{h}_v^{(+)} D_i^4 h_v^{(+)} ,
\end{equation}
written in the rest frame, $\bbox{v}=\bbox{0}$.
In the usual HQET, rotational invariance of continuum QCD implies
$w_4=0$.
With lattice QCD, however, $w_4$ does not vanish unless the lattice
action has been improved accordingly.

To describe electroweak transitions among hadrons containing a single 
heavy quark, HQET introduces effective operators for the interactions 
mediating the transitions.
Even in simple cases, such as the vector and axial vector currents
examined below, the number of operators in the heavy-quark expansion
is large, and the details of the construction are different for 
heavy-to-heavy and heavy-light transitions.
The notation for currents is postponed, therefore, to
Secs.~\ref{sec:semi} and~\ref{sec:fB}.

\section{Properties of~${\cal L}^{(0)}$}
\label{sec:properties}

The previous two sections establish that the heavy-quark limit of 
lattice QCD can be described by the effective 
Lagrangian~${\cal L}^{(0)}$, with small corrections from
\begin{equation}
	{\cal L}_I = {\cal L}^{(1)} + {\cal L}^{(2)} + \cdots.
	\label{eq:LI}
\end{equation}
This means that the eigenstates of lattice QCD are not very different 
from the eigenstates of the quantum field theory defined by
${\cal L}_{\text{light}}+{\cal L}^{(0)}$, where 
${\cal L}_{\text{light}}$ is the (Symanzik effective) Lagrangian of
the light quarks and gluons.
Apart from the rest mass~$m_1$ and lattice artifacts
of~${\cal L}_{\text{light}}$, this is the same lowest-order Lagrangian
that is used to describe heavy quarks in continuum~QCD.

To use HQET to connect lattice QCD to continuum QCD, one must 
understand how the rest mass and the higher-dimension interactions 
influence observables.
This section shows that the eigenstates of the Hamiltonian 
corresponding to ${\cal L}_{\text{light}}+{\cal L}_{\text{HQET}}$ 
are independent of the rest mass.
In particular, the eigenstates of
${\cal L}_{\text{light}}+{\cal L}^{(0)}$ 
do not depend on the heavy flavor at all.
The remainder of the paper then develops perturbation theory 
in~${\cal L}_I$ around these flavor-independent states and studies how 
the perturbations affect several observables.

To show that the rest mass~$m_1$ decouples from non-perturbative 
observables, it is convenient to switch to the Hamiltonian formalism 
of HQET.
The canonical conjugate to the field~$h_v$ is~\cite{Dugan:1992ak}
\begin{equation}
	\pi_v=iv^0\bar{h}_v
\end{equation}
so at equal times ($x^0=z^0$)
\begin{equation}
	\{h_v(x),\pi_v(z)\}=\{h_v(x),iv^0\bar{h}_v(z)\} =
		i\delta^{(3)}(\bbox{x}-\bbox{z})P_+(v).
	\label{eq:ccr}
\end{equation}
The Hamiltonian $H=\int d^3x\,{\cal H}$ has the density
\begin{equation}
	{\cal H} = {\cal H}_{\text{light}} + {\cal H}^{(0)} - {\cal L}_I,
	\label{eq:Hden}
\end{equation}
including a term for the light degrees of freedom.
The leading heavy-quark Hamiltonian density
\begin{eqnarray}
	{\cal H}^{(0)} & = & \pi_v\partial_0h_v - {\cal L}^{(0)} \\
		& = & m_1\bar{h}_vh_v + iv^0\bar{h}_vA^0h_v 
			- i\bar{h}_v\bbox{v}\cdot\bbox{D}h_v .
	\label{eq:H0}
\end{eqnarray}
From Eq.~(\ref{eq:ccr}) one can see that $\int d^3x\,\bar{h}_vh_v$
commutes with all other terms in~$H$, including with $H_{\text{light}}$
and $L_I=\int d^3x\,{\cal L}_I$.
Thus, the eigenstates of~$H$ are independent of~$m_1$.
This result is well known in other approaches to the heavy-quark 
limit~\cite{Lepage:1987gg,Uraltsev:1998bk}, but the general proof 
within HQET does not seem to be widely appreciated.%
\footnote{In specific examples, a small rest mass, called a 
\emph{residual mass}, has been shown to drop out of the $1/m_Q$ 
corrections~\cite{Falk:1992fm}.}

This result has a very important consequence.
In the HQET description of lattice QCD, lattice-spacing dependence 
appears in three places: the rest mass, the short-distance
coefficients of~${\cal L}_I$, and the light degrees of freedom.
Because the rest mass drops out of physical observables, it is 
acceptable---perhaps even advisable---to tolerate a discrepancy of the 
rest mass from the physical mass.
Genuine lattice artifacts of the heavy quark stem from deviations of
the \emph{higher}-dimension short-distance coefficients from their
continuum limit, and the couplings of the lattice action should be
tuned to minimize them.
To make this more point concrete, the effects of~${\cal L}_I$ can be
propagated to observables with tools developed for the usual HQET,
as shown in the rest of this paper.

\section{Perturbation Theory in ${\cal L}_I$}
\label{sec:ptLI}

The previous sections have established that lattice gauge theory with
heavy quarks can be described by the effective
Lagrangian~${\cal L}_{\text{HQET}}$, whose eigenstates are close to
those of the leading-order theory with Lagrangian~${\cal L}^{(0)}$.
To trace the effects of the higher-dimension operators in~${\cal L}_I$
on observables, they can be treated as perturbations.
A~formalism for perturbation theory that exploits heavy-quark
symmetry is reviewed in this section.

When proceeding to second order in~${\cal L}_I$, as in
Secs.~\ref{sec:masses} and~\ref{sec:semi} below, one must be careful
to be consistent, for example about the normalization of states.
Thus, the discussion starts (Sec.~\ref{sec:pt}) with a careful setup 
of time-ordered perturbation theory, to generate heavy-quark 
expansions based on the eigenstates of~${\cal L}^{(0)}$.
These states are desirable not only because they form mass-independent 
multiplets under heavy-quark symmetry, but also because they are 
affected by the lattice only through the light degrees of freedom.
The formalism makes no explicit reference to the short-distance
coefficients of the modified HQET, so it applies equally well to the
usual HQET and could be used there as well.
The heavy-quark expansion becomes a series of terms consisting of
short-distance coefficients multiplying matrix elements of time-ordered
products in the eigenstates of~${\cal L}^{(0)}$.
Many relations among these matrix elements follow from heavy-quark
symmetry, and Sec.~\ref{sec:tf} reviews the trace formalism, a
technique for deriving such relations.

\subsection{Time-ordered perturbation theory}
\label{sec:pt}

The perturbative series can be generated by generalizing the
interaction picture for vacuum expectation values to transition
matrix elements.
There are three quantum field theories to consider: the underlying
theory [here lattice QCD with action~(\ref{eq:S kappa})];
the full HQET with Lagrangian~(\ref{eq:L});
and the leading HQET with Lagrangian~(\ref{eq:L0}).
The states treated here are hadrons with one heavy quark.
The (lattice) QCD state with a heavy quark of flavor~$b$ ($c$) is 
denoted $|B\rangle$ ($|D\rangle$).
The analogous full HQET state is denoted $|B_v\rangle$, where the
subscript labels the chosen velocity.
Finally, the infinite-mass states are denoted $|b_vJ;j\alpha\rangle$, 
where $b$ is the heavy flavor in the HQET with velocity~$v$, $J$~is 
the hadron's spin, $j$~is the spin of the light degrees of freedom, 
and $\alpha$ encompasses all other quantum numbers of the light 
degrees of freedom.
By heavy-quark flavor and spin symmetry, the spatial wave functions
of these states do not depend on $b$ or~$J$.

By the Gell-Mann--Low theorem~\cite{Gell-Mann:1951rw} the lowest-lying
(\emph{i.e.}, $\alpha=0$) spin-$J$ hadron is related to the
corresponding infinite-mass state by
\begin{equation}
	|B_v\rangle = \lim_{T\to \infty(1-i0^+)}
		\frac{Z_B^{1/2}\;U(0,\mp T)|b_vJ;j0\rangle}%
		{\langle b_vJ;j0|U(0,\mp T)|b_vJ;j0\rangle},
	\label{eq:GML}
\end{equation}
where $Z_B$ is a state renormalization factor and
\begin{equation}
	U(t,t_0) = T\,\exp\int_{t_0}^td^4x\,{\cal L}_I
	\label{eq:U}
\end{equation}
is the familiar interaction-picture propagator.
The state renormalization factor has the usual interpretation of the 
overlap between the unperturbed and the fully dressed states:
$Z_B^{1/2}=\langle b_vJ;j0|B_v\rangle$.

To derive the heavy-quark expansion without ambiguities stemming from 
the normalization of states, one should set up perturbation theory so 
that~$Z_B$ does not appear. 
For example, the energy of a fully dressed state can be 
written~\cite{Gremm:1997df}
\begin{equation}
	E=\frac{\langle b_vJ;j0|H|B_v\rangle}{\langle b_vJ;j0|B_v\rangle}
		= \frac{\langle b_vJ;j0|H(0)U(0,-T)|b_vJ;j0\rangle}%
		       {\langle b_vJ;j0|    U(0,-T)|b_vJ;j0\rangle},
	\label{eq:E}
\end{equation}
in which the normalization of states clearly cancels.
Similarly, matrix elements for flavor-changing transitions can be 
expressed
\begin{equation}
	\frac{\langle D_{v'}|T\, O_1\cdots O_n|B_{v} \rangle^{\;\;}}%
	{ \langle D_{v'}|D_{v'}\rangle^{1/2}\;
	\;\langle B_{v} |B_{v} \rangle^{1/2}} =
	\frac{\langle c_{v'}J';j0|T\, O_1\cdots O_n
	e^{\int\!d^4x\,{\cal L}_I}|b_vJ;j0\rangle}%
	{ \langle c_{v'}J';j0|T\,e^{\int\!d^4x\,{\cal L}_I}
	|c_{v'}J';j0\rangle^{1/2}
	\;\langle b_{v} J ;j0|T\,e^{\int\!d^4x\,{\cal L}_I}
	|b_{v} J ;j0\rangle^{1/2}},
	\label{eq:ip}
\end{equation}
where the upper (lower) sign of Eq.~(\ref{eq:GML}) is used for the 
initial (final) state, and products of $U(t_1,t_2)$ have been 
coalesced according to its well-known properties~\cite{Peskin:1995ev}.
The factors $Z_B^{1/2}$ and $Z_D^{1/2}$ are eliminated in favor of the 
denominators by taking the modulus of each side of Eq.~(\ref{eq:GML}).

The operators $O_j$ in Eq.~(\ref{eq:ip}) are operators of HQET.
In general an operator from the underlying theory is described by a 
sum of operators in HQET, \emph{cf.}\ Secs.~\ref{sec:semi}
and~\ref{sec:fB}.
On the left-hand side the operators have the time dependence
\begin{equation}
	O_j(t) = e^{i H t} O_j e^{-i H t}
	\label{eq:HeisenbergO}
\end{equation}
of the Heisenberg picture, whereas on the right-hand side they have 
the time dependence
\begin{equation}
	O_j(t) = e^{iH_0t} O_j e^{-iH_0t}
	\label{eq:interactionO}
\end{equation}
of the interaction picture.
They are related by $O^{(H)}(t)=U^\dagger(t,0)O^{(I)}(t)U(t,0)$.
When~$O$ contains explicit time derivatives, as in some cases in
Appendix~\ref{app:traces}, the time dependence of the $U$s generates 
additional contact terms in the $T$-product in the interaction 
picture.

This setup of time-ordered perturbation theory is equivalent to 
Rayleigh-Schr\"odinger perturbation theory~\cite{Gell-Mann:1951rw}.
The denominators on the right-hand side of Eq.~(\ref{eq:ip}) are 
rarely made explicit in the literature on HQET, but they are 
necessary.
Indeed, in tracing the equivalence to Rayleigh-Schr\"odinger 
perturbation theory, one sees that the denominators generate 
wave-function renormalization and remove $|b_vJ;j0\rangle$ and 
$|c_{v'}J';j0\rangle$ from sums over intermediate states.
The procedure is analogous to taking connected vacuum correlation
functions.%
\footnote{Reference~\cite{Bigi:1995ga} notes both the significance
of the subtractions and the analogy with connected vacuum amplitudes,
but prefers not to use HQET.}
The infinitesimal in the limit $T\to\infty(1-i0^+)$ is needed to 
dampen the integrals in Minkowski space,
and it is unnecessary in Euclidean space.
There is no issue of analytic continuation here: the symbol~$U(t,t_0)$
is just an integral representation of the energy denominators in
ordinary perturbation theory.

The principal advantage of Eqs.~(\ref{eq:E}) and~(\ref{eq:ip}) is 
that they separate cleanly how each term in the heavy-quark expansion 
affects the matrix element on the left-hand side.
As desired, the normalization conditions on the full and 
infinite-mass states cancel separately.
Each operator in~${\cal L}_I$ can be treated one insertion after 
another, and the expansion leads to matrix elements in the 
mass-independent states $|b_vJ;j0\rangle$ and $\langle c_{v'}J';j0|$.
On the other hand, a formalism that starts with \emph{vacuum} 
expectation values of time-ordered products and proceeds to the 
left-hand side via the reduction formula leads to expressions with 
``in'' and ``out'' states whose masses equal those of the fully 
dressed states.

When employing HQET to describe lattice QCD it is especially helpful 
to obtain a series in the mass-independent eigenstates 
of~${\cal L}^{(0)}+{\cal L}_{\text{light}}$.
These states depend only mildly on the lattice spacing, through the 
light degrees of freedom.
Thus, the discretization effects of the heavy quark are truly
encapsulated into the short-distance coefficients of~${\cal L}_I$,
and one can estimate their effect simply by comparing the heavy-quark
expansions of continuum and lattice QCD.
With the expansions derived in subsequent sections, the comparison is
made easily by substituting the usual coefficients for the
modified~ones.

Although not strictly necessary, it is convenient to choose normalization 
conditions for the states.
In the underlying theory we normalize plane-wave states so that
\begin{equation}
	\langle B(\bbox{p}')|B(\bbox{p})\rangle =
		v^0(2\pi)^3\delta(\bbox{p}'-\bbox{p}),
	\label{eq:QCDstates}
\end{equation}
where $v^0=\sqrt{1+\bbox{v}^2}$.
In continuum QCD $\bbox{v}=\bbox{p}/M$ is the physical velocity of the 
true hadron, and in lattice QCD the relation between $\bbox{v}$ 
and~$\bbox{p}$ should tend to the same as $\bbox{p}a\to 0$.
Equation~(\ref{eq:QCDstates}) is convenient because it is 
relativistically invariant and its infinite mass limit is well 
behaved.
We also normalize full HQET states so that
\begin{equation}
	\langle B_v(\bbox{k}')|B_v(\bbox{k})\rangle =
		v^0(2\pi)^3\delta(\bbox{k}'-\bbox{k}),
	\label{eq:normstates}
\end{equation}
where $\bbox{k}^{(\prime)}$ is a small residual 
momentum~\cite{Dugan:1992ak}, 
and likewise for the infinite-mass states.
Note that in Eq.~(\ref{eq:normstates}) the factor of~$v^0$ does not 
introduce mass dependence; in HQET the velocity is an ingredient in 
the construction of the effective Lagrangian, not a property of the 
states.

To regulate $\delta$ functions one should smear plane-wave states into
wave packets before expanding out Eq.~(\ref{eq:E}) or~(\ref{eq:ip}).
With the same normalization condition~(\ref{eq:normstates}) for fully 
dressed and infinite-mass HQET states, the factors of~$v^0$ and the 
smearing functions cancel completely after expanding.
One can thus re-write Eq.~(\ref{eq:ip})
\begin{equation}
	\langle D_{v'}|T\, O_1\cdots O_n|B_v\rangle =
	\langle c_{v'}J';j0|T\, O_1\cdots O_n
	e^{\int\!d^4x\,{\cal L}_I}|b_vJ;j0\rangle^\star
	\label{eq:star}
\end{equation}
where the star on the right-hand side is a reminder to include 
the extra terms generated by expanding out the denominator of 
Eq.~(\ref{eq:ip}).
In Sec.~\ref{sec:semi} and Appendix~\ref{app:traces} this notation is 
used for $T$-products
$\langle c_{v'}J'|T\,O{\cal O}^b_X|b_vJ\rangle^\star$,
$\langle c_{v'}J'|T\,{\cal O}^c_XO{\cal O}^b_Y|b_vJ\rangle^\star$,
etc., where the operators~${\cal O}^h_X$ are those appearing 
in~${\cal L}_I$ for flavor~$h$.
The star means to collect all terms from the expansion with the
specified insertions.

\subsection{Trace formalism}
\label{sec:tf}

To evaluate the right-hand side of Eq.~(\ref{eq:star}) there is a 
powerful formalism, called the trace formalism, which takes full 
advantage of heavy-quark symmetry~\cite{Luke:1990eg}.
The objective is to calculate transition amplitudes of the form
\begin{equation}
	{\cal T}_{b_v\to c_{v'}}^{A_1\cdots A_N} = 
	\langle c_{v'}J';j0|T\, \bar{h}_{v'}\Gamma_1G_1^{A_1}h_{v'} \cdots
	\bar{h}_{v'}\Gamma_nG_n^{A_n}h_v \cdots
	\bar{h}_v\Gamma_NG_N^{A_N}h_v |b_vJ;j0\rangle^\star
	\label{eq:H'TH}
\end{equation}
and
\begin{equation}
	{\cal T}_{b_v\to 0}^{A_1\cdots A_N} = 
	\langle 0|T\, \bar{q}\Gamma_1G_1^{A_1}h_v\, 
	\bar{h}_v\Gamma_2G_2^{A_2}h_v \cdots
	\bar{h}_v\Gamma_NG_N^{A_N}h_v |b_vJ;j0\rangle^\star,
	\label{eq:0TH}
\end{equation}
where the $G_k^{A_k}$ is a combination of covariant
derivatives~${\cal D}$ (including field strengths~$F^{\mu\nu}$) 
and light-quark bilinears~$\bar{q}q$ with
Lorentz indices abbreviated by the superscript~$A_k$.

The color and spin dependence of each static propagator
$Th_v(x)\bar{h}_v(y)$ [or $Th_{v'}(x)\bar{h}_{v'}(y)$] factors into
a Wilson line and a projector $P_+(v)=:P_+$ [or $P_+(v')=:P'_+$].
That means that the amplitudes can be written
(for $j=\case{1}{2}$ mesons)
\begin{equation}
	{\cal T}_{b_v\to c_{v'}}^{A_1\cdots A_N} = 
	-\tr\{\bar{\cal M}_{J'}(v')\Gamma_1P'_+ \cdots P'_+\Gamma_n P_+
	\cdots P_+\Gamma_N {\cal M}_J(v) \Xi^{A_1\cdots A_N} \},
\end{equation}
and
\begin{equation}
	{\cal T}_{b_v\to 0}^{A_1\cdots A_N} = 
	- \tr\{\Gamma_1P_+\Gamma_2P_+\cdots P_+\Gamma_N {\cal M}_J(v) 
		\Xi^{A_1\cdots A_N} \},
\end{equation}
where ${\cal M}_J$ and $\bar{\cal M}_{J'}$ are spin wave functions and 
$\Xi^{A_1\cdots A_N}$ parametrizes the spatial wave functions and a 
trace over color of the Wilson lines, punctuated by the $G^{A_k}$, 
with the light quark propagator.
There is only one trace over heavy-quark spin, because products of 
traces correspond to disconnected terms, which are subtracted when 
expanding Eq.~(\ref{eq:star}).
The minus sign arises because the trace over spin is obtained after
anti-commuting the left-most quark field all the way to the right.

Spin wave functions such as ${\cal M}_J(v)$ and~$\bar{\cal M}_{J'}(v')$
are determined by spin symmetry alone.
For $j=\case{1}{2}$ they are
\begin{eqnarray}
	{\cal M}_0(v) & = & i2^{-1/2} P_+(v)\gamma_5, \\
	{\cal M}_1(v) & = & i2^{-1/2} P_+(v)
		\kern+0.1em /\kern-0.55em \epsilon.
\end{eqnarray}
Charge conjugates are $\bar{\cal M}=\gamma_4{\cal M}^\dagger\gamma_4$,
\begin{eqnarray}
	\bar{\cal M}_0(v) & = & i2^{-1/2}\gamma_5 P_+(v), \\
	\bar{\cal M}_1(v) & = & i2^{-1/2}
		\kern+0.1em /\kern-0.55em \bar{\epsilon}\,P_+(v),
\end{eqnarray}
where $\bar{\epsilon}=\epsilon^{*}$ in Minkowski space-time
and $\bar{\epsilon}=(\bbox{\epsilon}^{*}, -\epsilon_4^*)$ in Euclidean
space-time. 
Note that ${\cal M}=P_+{\cal M}P_-$
and $\bar{\cal M}=P_-\bar{\cal M}P_+$.
Generalizations to $j=0$ and $j=1$ 
baryons~\cite{Georgi:1991cx,Mannel:1991vg} and to higher angular 
momentum~\cite{Falk:1992nq} are available in the literature.

The functions $\Xi^{A_1\cdots A_N}$ cannot be obtained from symmetry 
considerations alone.
They depend on the velocities $v'$ and $v$ and the quantum numbers of 
the light degrees of freedom.
They parametrize the long-distance dynamics of 
${\cal L}^{(0)}+{\cal L}_{\text{light}}$, so they do not depend on 
flavor, and they suffer from lattice artifacts only of the light 
degrees of freedom.
As explained above, cutoff effects of the heavy quark are captured in 
the coefficients of the modified HQET, which multiply matrix 
elements~(\ref{eq:H'TH}) and~(\ref{eq:0TH}).

\section{Hadron Masses}
\label{sec:masses}

The simplest application of the HQET formalism is to generate an
expansion for the rest mass of a heavy-light hadron.
In numerical lattice calculations the energy of a state of 
momentum~$\bbox{p}$ is computed by looking at the (imaginary) time 
evolution of a correlation function
\begin{equation}
\langle\Phi_{\bbox{p}'}(x_4)\Phi_{\bbox{p}}^\dagger(0)\rangle =
\delta_{\bbox{p}'\bbox{p}}  \left[
  \theta(x_4) \sum_{n}  e^{-x_4E_{n}(\bbox{p})}
     |\langle B_n |\Phi_{\bbox{p}}^\dagger|0\rangle|^2
+ \theta(-x_4)\sum_{n'} e^{x_4E_{n'}(\bbox{p})}
     |\langle \bar{B}_{n'}|\Phi_{\bbox{p}}|0\rangle|^2\right],
\end{equation}
where $|B_n\rangle$ ($|\bar{B}_{n'}\rangle$) are full lattice-QCD
states connected to the vacuum by $\Phi_{\bbox{p}}^\dagger$
($\Phi_{\bbox{p}}$).
By a combination of judicious choices of $\Phi_{\bbox{p}}^\dagger$ and 
taking~$x_4$ large enough, one can isolate the lower-lying states.
At small momentum, the relation between energy and momentum is
\begin{equation}
E(\bbox{p})=M_1 +\frac{\bbox{p}^2}{2M_2},
\end{equation}
which defines the hadron's rest mass~$M_1$ and kinetic mass $M_2$.
(Some authors call~$M_1$ the ``pole'' mass, but~$M_1$ and~$M_2$
are both properties of the particle's pole.)
In this paper upper-case is used to denote hadron masses,
and lower-case to denote quark masses.

These energies can be thought of as eigenvalues of a Hamiltonian,
defined via the transfer matrix,
which HQET models with Eq.~(\ref{eq:Hden}).
In Eq.~(\ref{eq:E}) $H$ is always to the left of~$U(0,-T)$, so one
can make the split $H=H_{\text{light}}+H^{(0)}-L_I$ and act the first
two terms on the bra~$\langle b_vJ;j0|$.
Setting~$\bbox{p}=\bbox{0}$ and calling the leading eigenvalue 
\begin{equation}
	m_1 + \bar{\Lambda} = \frac{%
		\langle b_vJ;j0|[H_{\text{light}}+H^{(0)}]|b_vJ;j0\rangle}{%
		\langle b_vJ;j0|b_vJ;j0\rangle} ,
	\label{eq:Lambda}
\end{equation}
the heavy-quark expansion of the hadron mass is generated by
\begin{equation}
	M_1 = m_1 + \bar{\Lambda} -
	\langle b_vJ;j0|L_I\,T\,
	e^{\int\!d^4x\,{\cal L}_I}|b_vJ;j0\rangle^\star,
	\label{eq:M1exp}
\end{equation}
where $L_I$ is at time~$0$, the time integration is from $-\infty$ 
to~$0$, and the star is a reminder not to neglect the denominator
in Eq.~(\ref{eq:E}).
The quark's rest mass enters solely additively because its term in the 
Hamiltonian commutes with all others.

The expansion of Eq.~(\ref{eq:M1exp}) leads to reduced matrix elements
that depend on the spin~$j$ of the light degrees of freedom ($j=0$
for the $\Lambda_b$ baryons, $j=1/2$ for the $B$ and $B^*$ mesons,
etc.), but not on the heavy quark's spin.
Through order $1/m_Q^2$ one defines
\begin{eqnarray}
	\langle b_vJ;j0|{\cal O}_2|b_vJ;j0\rangle & = & \lambda_1,
	\label{eq:lambda1} \\
	\langle b_vJ;j0|{\cal O}_B|b_vJ;j0\rangle & = & d_J\lambda_2,
	\label{eq:lambda2} \\
	\langle b_vJ;j0|{\cal O}_D|b_vJ;j0\rangle & = & -2\rho_1,
	\label{eq:rho1} \\
	\langle b_vJ;j0|{\cal O}_E|b_vJ;j0\rangle & = & -2d_J\rho_2,
	\label{eq:rho2}
\end{eqnarray}
and, in the notation of Ref.~\cite{Gremm:1997df},
\begin{eqnarray}
	\int d^4x\,\langle b_vJ;j0|{\cal O}_2(0)
	{\cal O}_2(x)|b_vJ;j0\rangle^\star & = & {\cal T}_1,
	\label{eq:T1} \\
	\int d^4x\,\langle b_vJ;j0|{\cal O}_B(0)
	{\cal O}_B(x)|b_vJ;j0\rangle^\star & = & 
		{\cal T}_3 + d_J({\cal T}_4 - {\cal T}_2),
	\label{eq:T3} 
\end{eqnarray}
\begin{equation}
	\int d^4x\,\langle b_vJ;j0|{\cal O}_2(0)
		{\cal O}_B(x)|b_vJ;j0\rangle^\star =
	\int d^4x\,\langle b_vJ;j0|{\cal O}_B(0)
		{\cal O}_2(x)|b_vJ;j0\rangle^\star = d_J{\cal T}_2.
	\label{eq:T2}
\end{equation}
The $J$-dependence in Eqs.~(\ref{eq:lambda1})--(\ref{eq:T2}) is $d_0=3$
(for the $B$ meson) and $d_1=-1$ (for the $B^*$ meson).
For the $\Lambda_b$ baryon there are fewer non-vanishing matrix 
elements; the above formulae hold if one sets $d_{1/2}=0$.
The parameters~$\bar{\Lambda}$, $\lambda_n$, $\rho_n$, and~${\cal T}_n$
are the same as in continuum QCD, apart from lattice artifacts of the
light degrees of freedom.
Combining Eqs.~(\ref{eq:M1exp})--(\ref{eq:T2}) the rest mass becomes
\begin{equation}
 \!	M_1 = m_1 + \bar{\Lambda}
		- \frac{\lambda_1}{2m_2} - \frac{d_J\lambda_2}{2m_B}
		+ \frac{\rho_1}{4m_D^2}  + \frac{d_J\rho_2}{4m_E^2}
		- \frac{{\cal T}_1}{4m^2_2} 
		- \frac{2d_J{\cal T}_2}{2m_2\,2m_B}
		- \frac{{\cal T}_3 + d_J({\cal T}_4-{\cal T}_2)}{4m^2_B}.
	\label{eq:M1}
\end{equation}
The result~(\ref{eq:M1}) is simple enough that it could have been 
written down upon inspection of Eqs.~(\ref{eq:L1}) and~(\ref{eq:L2}) 
and comparing to the continuum 
papers~\cite{Falk:1993wt,Mannel:1994kv,Bigi:1995ga,Gremm:1997df}.

This result is the first example of the expansion for which
Eq.~(\ref{eq:schematic}) is a prototype.
Short-distance effects of the heavy quark, including lattice-spacing 
effects, are contained in the ``masses''~$m_1$, $m_2$, $m_B$, $m_D$, 
and~$m_E$.
If the bare mass is adjusted so that $m_2=m_Q$, then the mass
formula~(\ref{eq:M1}) shows that the spin-averaged splittings, such
as $m_{\Lambda_b}-\case{1}{4}(m_B+3m_{B^*})$, are reproduced correctly
to order~$1/m_Q$.
The Sheikholeslami-Wohlert action has a second parameter, with
which~$1/m_B$ can be adjusted (via a short-distance calculation)
to reproduce correctly the spin splittings, such as $m_{B^*}-m_B$,
to order~$1/m_Q$.
These adjustments are essential, because in matrix elements the rest
mass plays no role whatsoever.

In the usual HQET with $m_1=0$, the quark mass is added
to~$\bar{\Lambda}$ and the higher-order terms.
Ambiguities of the HQET renormalization scheme, including those of 
infrared renormalons in the on-shell scheme, cancel in the sum.
Similarly, the difference $m_2-m_1$ can be added to Eq.~(\ref{eq:M1}):
$M=M_1+m_2-m_1$.
Adding the residual mass in this way has the virtue that $m_2-m_1$
does not suffer from infrared ambiguities, even in the on-shell scheme.

\section{Semileptonic Form Factors}
\label{sec:semi}

Another interesting application of HQET is the heavy-quark expansion
of form factors in the exclusive semileptonic decays $B\to D^*l\nu$
and $B\to D l\nu$.
These decays offer the most promising way to decrease the uncertainty 
in the CKM element~$|V_{cb}|$, provided the hadronic matrix elements 
can be calculated reliably.
Recent work~\cite{Hashimoto:1999yp,Simone:1999nv} shows that
calculations of the form factors at zero recoil with statistical
errors of a few percent are feasible.
The aim of this section is to describe the~$1/m_Q$ and~$1/m_Q^2$
contributions to the lattice observables calculated in
Refs.~\cite{Hashimoto:1999yp,Simone:1999nv}, and compare them to the
description of the form factors in the usual~HQET.
The technical details are in Appendix~\ref{app:traces}, mostly
following Refs.~\cite{Falk:1993wt,Mannel:1994kv}.

The transitions are mediated by the charged weak currents
\begin{equation}
	{\cal V}^\mu = \bar{c}i\gamma^\mu b ,\quad
	\label{eq:vector-current} \\
	{\cal A}^\mu = \bar{c}i\gamma^\mu\gamma_5 b ,
	\label{eq:axial-vector-current}
\end{equation}
where $\bar{c}$ and $b$ are conventionally normalized continuum 
quark fields.
Currents in lattice gauge theory and in HQET are introduced below,
but the symbols~${\cal V}^\mu$ and~${\cal A}^\mu$ are reserved for
the physical currents.
The hadronic part of the transitions involves the matrix elements
$\langle D^{(*)}|{\cal V}^\mu|B\rangle$ and
$\langle D^*|{\cal A}^\mu|B\rangle$.
For $B\to D l\nu$ there are two form factors $h_+$ and $h_-$.
With the normalization~(\ref{eq:normstates}) they are related to the 
matrix element by
\begin{equation}
	\langle D(\bbox{v}')|{\cal V}^\mu|B(\bbox{v})\rangle =
		\case{1}{2}(v'+v)^\mu h_+(w) - \case{1}{2}(v'-v)^\mu h_-(w),
	\label{eq:h+h-}
\end{equation}
where $w=-v'\cdot v$.
Zero recoil corresponds to~$w=1$.
In Eq.~(\ref{eq:h+h-}) the final velocity is kept distinct from the 
initial velocity to be able to obtain~$h_-(1)$.
For $B\to D^*l\nu$ there are three axial form factors, defined by
\begin{equation}
	\langle D^*(\bbox{v}',\epsilon')|{\cal A}^\mu|B(\bbox{v})\rangle =
	\case{1}{2}(w+1)i\overline{\epsilon'}^\mu h_{A_1}(w) +
	\case{1}{2}i\overline{\epsilon'}\!\cdot\!v\,v^\mu h_{A_2}(w) +
	\case{1}{2}i\overline{\epsilon'}\!\cdot\!v\,v^{\prime\mu} h_{A_3}(w),
	\label{eq:hA1}
\end{equation}
and a vector form factor, but at zero recoil the decay rate depends
only on~$h_{A_1}(1)$.
For reasons that will become clear below, the zero-recoil matrix element
\begin{equation}
	\langle D^*(\bbox{v},\epsilon')|{\cal V}^\mu
	|B^*(\bbox{v},\epsilon)\rangle = 
		\overline{\epsilon'}\!\cdot\!\epsilon\,v ^\mu h_1(1)
	\label{eq:h1}
\end{equation}
and its form factor~$h_1(1)$ are also of interest.

Note that continuum QCD currents define the form factors.
To generate the heavy-quark expansion of these form factors,
one replaces the currents~${\cal V}^\mu$ and~${\cal A}^\mu$
with effective currents built from the heavy-quark fields and the
fields of the light degrees of freedom.
The effective currents and the heavy-quark Lagrangian are treated to 
the desired order in~$1/m_Q$, and Eq.~(\ref{eq:ip}) should be used to 
generate the expansion, consistent to that order.

The zeroth order is simple and worth reviewing briefly.
The QCD currents are related to HQET currents via
\begin{eqnarray}
	{\cal V}^\mu & \doteq &
		\eta_V\bar{c}_{v'}i\gamma^\mu b_v -
		\case{1}{2}\beta_V  (v'-v)^\mu \bar{c}_{v'} b_v -
		\case{1}{2}\gamma_V (v'-v)_\nu \bar{c}_{v'}i\sigma^{\mu\nu}b_v,
	\label{eq:calV=effV} \\
	{\cal A}^\mu & \doteq &
		\eta_A\bar{c}_{v'}i\gamma^\mu\gamma_5 b_v -
		\case{1}{2}\beta_A  (v'-v)^\mu \bar{c}_{v'}\gamma_5 b_v -
		\case{1}{2}\gamma_A (v'-v)_\nu
		\bar{c}_{v'}i\sigma^{\mu\nu}\gamma_5 b_v,
	\label{eq:calA=effA}
\end{eqnarray}
where the symbol~$\doteq$ means that the operators, though defined
in different field theories, have the same matrix elements.
The short-distance coefficients depend on the two masses;
$\eta_j$ and $\gamma_j$ are symmetric upon interchanging the masses
($j\in\{V,A\}$); $\beta_j$ is anti-symmetric; at the tree level they
satisfy~$\eta_j=1$, $\beta_j=\gamma_j=0$.
To obtain the leading heavy-quark expansion, one simply takes matrix 
elements of the effective currents in the states of the infinite-mass 
theory.
From the trace formalism one finds
\begin{eqnarray}
	h_+(w) & = &
	\left[\eta_V + \case{1}{2}(w-1)\gamma_V\right]\xi(w) + O(1/m_Q),
	\label{eq:h+O(1)} \\
	h_-(w) & = & \case{1}{2}(w+1)\beta_V\xi(w) + O(1/m_Q),
	\label{eq:h-O(1)} \\
	h_1(w) & = &
	\left[\eta_V + \case{1}{2}(w-1)\gamma_V\right]\xi(w) + O(1/m_Q),
	\label{eq:h1O(1)} \\
	h_{A_1}(w) & = & \eta_A \xi(w) + O(1/m_Q), \label{eq:hA1O(1)}
\end{eqnarray}
with a single HQET form factor~$\xi(w)$, called the Isgur-Wise
function.
At zero recoil it is normalized by heavy-quark
symmetry~\cite{Isgur:1989vq}, so $\xi(1)=1$.
Therefore, the leading term in heavy-quark expansion is
$h_+(1)=h_1(1)=\eta_V$, $h_-(1)=\beta_V$, and $h_{A_1}(1)=\eta_A$.

The $1/m_Q$~\cite{Luke:1990eg} and
$1/m_Q^2$~\cite{Falk:1993wt,Mannel:1994kv} corrections to
Eqs.~(\ref{eq:h+O(1)})--(\ref{eq:hA1O(1)}) have been worked out
with~HQET.
This section repeats the analysis through order~$1/m_Q^2$ for the 
lattice approximants to the form factors introduced in 
Refs.~\cite{Hashimoto:1999yp,Simone:1999nv}.
The only crucial difference is that the short-distance coefficients 
are tracked carefully and their contributions are kept separate in the 
final results.

\subsection{Lattice and HQET currents}
\label{sec:currents}

To compute the form factors in 
Eqs.~(\ref{eq:h+h-})--(\ref{eq:h1}) with lattice gauge theory one 
introduces combinations of lattice fields with the same quantum 
numbers as~${\cal V}^\mu$ and~${\cal A}^\mu$.
The lattice currents are given by a series of dimension-three, -four, 
-five, etc., operators, with coefficients chosen to attain the right 
normalization and to reduce lattice artifacts.
Several choices have been made in the literature, but with Wilson
fermions they can all be described by HQET: the different choices
simply have different short-distance coefficients.

Let~$Z_{V^{cb}}V^\mu_{\text{lat}}$ ($Z_{A^{cb}}A^\mu_{\text{lat}}$)
denote the lattice approximant to the charged $b\to c$ vector
(axial-vector) current.
To conform with much of the literature on lattice gauge theory, the 
current's normalization factor~$Z_{j^{cb}}$ in shown explicitly.
Then, suppressing the space-time index, the lattice currents are 
related to HQET currents via
\begin{eqnarray}
	Z_{V^{cb}}V_{\text{lat}} & \doteq &
		V^{(0)} + \sum_{s=1} \sum_{r=0}^s V^{(r,s-r)} \\
	               & \doteq &
		V^{(0)}   + V^{(0,1)} + V^{(1,0)} +
		V^{(0,2)} + V^{(1,1)} + V^{(2,0)} + \cdots
\end{eqnarray}
and similarly for~$A^\mu_{\text{lat}}$.
The HQET operator~$V^{(r,s)}$ carries dimension $3+r+s$.
To make contact with the usual HQET, it is helpful to think of the 
dimensions being balanced by $r$~powers of~$1/m_c$ and $s$~powers
of~$1/m_b$.
The dimension-three vector current is
\begin{equation}
	V_\mu^{(0)} = 
		(\eta_V+\delta\eta_{V_\mu}^{\text{lat}})
		\bar{c}_{v'}i\gamma_\mu b_v -
		\case{1}{2}\beta_{V_\mu}^{\text{lat}} (v'-v)_\mu \bar{c}_{v'} b_v -
		\case{1}{2}\gamma_{V_{\mu\nu}}^{\text{lat}} 
		(v'-v)^\nu \bar{c}_{v'}i\sigma_{\mu\nu}b_v .
	\label{eq:V0}
\end{equation}
In general the coefficients depend on the directional indices,
because the lattice singles out the time direction.
The overall factor~$Z_V$ is conventionally chosen so that 
$\delta\eta_{V_0}^{\text{lat}}=0$.
Then $\delta\eta_{V_i}^{\text{lat}}$ vanishes at the tree level, but not
in general.
As with the usual HQET $\beta_{V_\mu}^{\text{lat}}$ and
$\gamma_{V_{\mu\nu}}^{\text{lat}}$ are, respectively, antisymmetric
and symmetric upon interchange of heavy quark masses and both vanish
at the tree level.
The operator multiplying $\beta_{V_\mu}^{\text{lat}}$
($\gamma_{V_{\mu\nu}}^{\text{lat}}$) makes a contribution at first
(second) order in~$v'-v$.

At dimension four and higher many operators arise, and a complete
catalog requires a voluminous notation.
Only the $\eta$-like terms are listed here.
$\beta$-like terms are not needed until Sec.~\ref{sec:near}, 
and $\gamma$-like terms are not needed at all.
With this restriction, the dimension-four currents are
\begin{eqnarray}
	V_\mu^{(0,1)} & = & -\eta_V^{(0,1)} \frac{%
	\bar{c}_{v'}i\gamma_\mu {\kern+0.1em /\kern-0.65em D}_\perp b_v}{%
	2m_{3b}} ,
	\label{eq:V01} \\
	V_\mu^{(1,0)} & = & +\eta_V^{(1,0)} \frac{\bar{c}_{v'}%
	\loarrow{\kern+0.1em /\kern-0.65em D}_{\perp'}i\gamma_\mu b_v}{%
	2m_{3c}},
	\label{eq:V10}
\end{eqnarray}
where $D_{\perp'}=D+v'\cdot Dv'$.
The notation $\eta_V^{(1,0)}/m_{3c}$ and $\eta_V^{(0,1)}/m_{3b}$
for the short-distance coefficients follows a helpful convention:
for degenerate quarks the coefficient is merely $1/m_3$, which thus
depends only on the indicated flavor; $\eta_V^{(r,s)}$ then describes
the additional radiative corrections for non-degenerate masses.
The dimension-five currents are
\begin{eqnarray}
	V_\mu^{(0,2)} & = & \eta_{VD^2_\perp}^{(0,2)}
	\frac{\bar{c}_{v'}i\gamma_\mu D^2_\perp b_v}{8m^2_{D^2_\perp b}}
	+ \eta_{VsB}^{(0,2)} \frac{%
	\bar{c}_{v'}i\gamma_\mu s^{\alpha\beta}B_{\alpha\beta} b_v}{%
	8m^2_{sB b}} 
	- \eta_{V\alpha E}^{(0,2)} \frac{%
	\bar{c}_{v'}i\gamma_\mu i{\kern+0.1em /\kern-0.65em E} b_v}{%
	4m^2_{\alpha E b}} , \label{eq:V02} \\
	V_\mu^{(2,0)} & = & \eta_{VD^2_\perp}^{(2,0)}
	\frac{\bar{c}_{v'}\loarrow{D}^2_{\perp'} 
	i\gamma_\mu b_v}{8m^2_{D^2_\perp c}}
	+ \eta_{VsB}^{(2,0)} \frac{%
	\bar{c}_{v'}s^{\alpha\beta}B'_{\alpha\beta}i\gamma_\mu b_v}{%
	8m^2_{sB c}} 
	+ \eta_{V\alpha E}^{(2,0)} \frac{%
	\bar{c}_{v'}i{\kern+0.1em /\kern-0.65em E}'i\gamma_\mu b_v}{%
	4m^2_{\alpha E c}} , \label{eq:V20} \\
	V_\mu^{(1,1)} & = &  - z^{(1,1)}_{V1} \frac{%
	\bar{c}_{v'}(\loarrow{\kern+0.1em /\kern-0.65em D}_{\perp'}
	i\gamma_\mu {\kern+0.1em /\kern-0.65em D}_\perp)_1b_v}{%
	2m_{3c}\,2m_{3b}} - z^{(1,1)}_{Vs} \frac{%
	\bar{c}_{v'}(\loarrow{\kern+0.1em /\kern-0.65em D}_{\perp'}
	i\gamma_\mu {\kern+0.1em /\kern-0.65em D}_\perp)_sb_v}{%
	2m_{3c}\,2m_{3b}} , \label{eq:V11}
\end{eqnarray}
where again $1/m^2_{Xh}$ depends only on the indicated flavor
and~$\eta_V^{(r,s)}$ depends on both masses.
The two coefficients~$z_{V1}^{(1,1)}$ and~$z_{Vs}^{(1,1)}$ multiply the
spin-independent and spin-dependent part of the Dirac matrix structure.
They do not reduce to~1 for equal masses, because $1/m_3$ is defined
through the dimension-four currents, but for most choices of the lattice
current they do equal~1 at the tree level.

\subsection{At zero recoil: $h_+(1)$ and $h_1(1)$}

The matrix elements that are to be described are
\begin{equation}
	\langle D| Z_jj_{\text{lat}} |B \rangle =
		\langle D_{v'}| j^{(0)} |B_{v} \rangle +
		\langle D_{v'}| j^{(1)} |B_{v} \rangle +
		\langle D_{v'}| j^{(2)} |B_{v} \rangle ,
	\label{eq:lat=hqet}
\end{equation}
where $j$ is $V$ or $A$, and $j^{(1)}=j^{(0,1)}+j^{(1,0)}$,
$j^{(2)}=j^{(0,2)}+j^{(1,1)}+j^{(2,0)}$.
The first two matrix elements on the right-hand side of 
Eq.~(\ref{eq:lat=hqet}) must be expanded via Eq.~(\ref{eq:ip}) to 
second and first order in~${\cal L}_I$, respectively.
There are, consequently, many HQET matrix elements to introduce.
The matrix elements and their abbreviations, analogous to those in
Sec.~\ref{sec:masses}, are listed in Table~\ref{tab:not}.
\begin{table}[t]
	\centering
	\caption[tab:not]{Notation for HQET matrix elements in
	Refs.\cite{Falk:1993wt,Mannel:1994kv} and this work.}
	\begin{tabular}{cccc}
		contribution & Ref.~\cite{Falk:1993wt} & 
			Ref.~\cite{Mannel:1994kv} & this work  \\
		\hline
		$\langle j^{(0)}\rangle$ & $\xi(w)$ & 1 & $\xi(w)$ \\
		$\langle j^{(1)}\rangle$ & $\xi_{\pm,\,3}(w)$ &   & 
			$\xi_{\pm,\,3}(w)$  \\
		$\langle Tj^{(0)}{\cal O}_2\rangle^\star$ & $A_1(w)$ & $\chi_1$ & 
			$A_1(w)$ \\
		$\langle Tj^{(0)}{\cal O}_B\rangle^\star$ & $A_{2,\,3}(w)$ & 
			$\chi_3$ & $A_{2,\,3}(w)$ \\
		$\langle j^{(2)}\rangle$ & $\phi_{0,\ldots,3}(w)$ & 
			$\lambda_{1,\,2}$ & $\lambda_{1,\ldots,4}(w)$  \\
		$\langle Tj^{(1)}{\cal O}_2\rangle^\star$ & 
			$E_{1,\,2,\,3}(w)$; $E'_{1,\,2,\,3}(w)$ &   & 
			$\Xi_{2,\,3}(w)$;   $F_{1,\,3}(w)$ \\
		$\langle Tj^{(1)}{\cal O}_B\rangle^\star$ & 
			$E_{4,\ldots,11}(w)$;   $E'_{4,\ldots,11}(w)$ &   & 
			$\Xi_{4,\ldots,11}(w)$; $F_{4,\ldots,11}(w)$ \\
		$\langle Tj^{(0)}{\cal Q}_D\rangle^\star$ & $2B_1(w)$ & $2\Xi_1$ & 
			$\lambda_1$ \\
		$\langle Tj^{(0)}{\cal Q}_E\rangle^\star$ & $2B_{2,\,3}(w)$ & 
			$2\Xi_3$ & $\lambda_2$ \\
		$\langle T{\cal O}_2j^{(0)}{\cal O}_2\rangle^\star$ & 
			$D_1(w)$ & $D$ & $D$ \\
		$\langle T{\cal O}_2j^{(0)}{\cal O}_B\rangle^\star$ & 
			$D_{2,\,3}(w)$ & $E$ & $E$ \\
		$\langle T{\cal O}_Bj^{(0)}{\cal O}_B\rangle^\star$ & 
			$D_{4,\ldots,10}(w)$ & $R_{1,\,2}$ & $R_{1,\,2}$ \\
		$\langle Tj^{(0)}{\cal O}_2{\cal O}_2\rangle^\star$ & 
			$C_1(w)$ & $A$ & $A=-\case{1}{2}D$ \\
		$\langle Tj^{(0)}{\cal O}_2{\cal O}_B\rangle^\star$ & 
			$C_{2,\,3}(w)$ & $B$ & $B=-E$ \\
		$\langle Tj^{(0)}{\cal O}_B{\cal O}_B\rangle^\star$ & 
			$C_{4,\ldots,12}(w)$ & $C_{1,\,3}$ & 
			$C_{1,\,3}=-\case{1}{2}R_{1,\,2}$ \\
	\end{tabular}
	\label{tab:not}
\end{table}
The notation mostly follows previous 
work~\cite{Falk:1993wt,Mannel:1994kv}.

One can work out the matrix elements using the trace formalism.
At zero recoil $\langle D_{v}| j^{(1)} |B_{v}\rangle$ vanishes.
For the vector-current transitions $B\to D$ and $B^*\to D^*$ with
$\bbox{v}=\bbox{v}'=\bbox{0}$ one finds
\begin{equation}
	\langle D^{(*)}|Z_{V^{cb}}V_{\text{lat}}^0|B^{(*)}\rangle =
		\eta_V W^{(0)}_{JJ} + W^{(2)}_{JJ},
	\label{eq:DVB}
\end{equation}
in which $\langle D_{v}|V^{(0)}|B_{v}\rangle$ yields
\begin{eqnarray}
	W^{(0)}_{JJ} = 1
	&+&	\left(\frac{1}{4m_{2c}^2}+\frac{1}{4m_{2b}^2}\right) A
	 +	\left(\frac{1}{2m_{2c}} \frac{1}{2m_{Bc}} +
		      \frac{1}{2m_{2b}} \frac{1}{2m_{Bb}}\right) d_J B
	\nonumber \\
	&+&	\left(\frac{1}{4m_{Bc}^2}+\frac{1}{4m_{Bb}^2}\right) [C_1+d_JC_3]
	 +	\frac{1}{2m_{2c}} \frac{1}{2m_{2b}} D
	\label{eq:W0JJ} \\
	&+&	\left(\frac{1}{2m_{Bc}} \frac{1}{2m_{2b}} +
	          \frac{1}{2m_{2c}} \frac{1}{2m_{Bb}}\right) d_J E 
	 +	\frac{1}{2m_{Bc}} \frac{1}{2m_{Bb}} [R_1+d_JR_2],
	\nonumber
\end{eqnarray}
and $\langle D_{v}|V^{(2)}|B_{v}\rangle$ yields
\begin{equation}
	W^{(2)}_{JJ} =
		\left(\frac{\eta_{VD_\perp^2}^{(2,0)}}{8m_{D_\perp^2c}^2}
		  +	  \frac{\eta_{VD_\perp^2}^{(0,2)}}{8m_{D_\perp^2b}^2}
		  -	  \frac{z^{(1,1)}_{V1}}{2m_{3c}\,2m_{3b}} \right) \lambda_1 
	  +	\left(\frac{\eta_{VsB}^{(2,0)}}{8m_{sBc}^2}
	  	  +	  \frac{\eta_{VsB}^{(0,2)}}{8m_{sBb}^2}
		  -   \frac{z^{(1,1)}_{Vs}}{2m_{3c}\,2m_{3b}} \right) d_J \lambda_2.
	\label{eq:W2JJ}
\end{equation}
The subscript $JJ'$ denotes the initial and final spins, although
here~$J'=J$.
The spin factor $d_0=3$ for $B\to D$ and $d_1=-1$ for $B^*\to D^*$.
The coefficient factors reveal the origin of the contribution.
By heavy-quark symmetry~$\lambda_1$ and $\lambda_2$ are exactly the
same as in Sec.~\ref{sec:masses}, and $A$, $B$, $C_1$, $C_3$, $D$,
$E$, $R_1$, and $R_2$ are new constants parametrizing the light
degrees of freedom, introduced in Appendix~\ref{app:traces} and 
the last six rows of Table~\ref{tab:not}.

Equation~(\ref{eq:DVB}) gives lattice approximants to the form
factors~$h_+(1)$ and~$h_1(1)$.
One striking feature of Eqs.~(\ref{eq:DVB})--(\ref{eq:W2JJ}) is that
there are no contributions of order~$1/m_Q$.
For continuum QCD, this is known as Luke's theorem~\cite{Luke:1990eg}.
Matrix elements of $j^{(1)}$ in the infinite-mass states contribute 
only when $v'\neq v$, so a single power of $1/m_3$ does not appear in 
Eq.~(\ref{eq:DVB}).
As shown in Appendix~\ref{app:luke}, terms with a single power 
of~$1/m_2$ and~$1/m_B$ are absent as a consequence of heavy-quark 
symmetry and Eq.~(\ref{eq:ip}).
Thus, Luke's theorem holds for lattice QCD also.

At order~$1/m_Q^2$, matrix elements that might have multiplied
$1/m_D^2$ and~$1/m_E^2$ also vanish; so do matrix elements
involving four-quark operators.
Furthermore, the parameters $A$, $B$, $C_1$, and $C_3$ can be
eliminated, as indicated in the right-most column of the last three
rows of Table~\ref{tab:not}.
As shown in Appendix~\ref{app:fluke}, this is another consequence of 
heavy-quark symmetry and Eq.~(\ref{eq:ip}).
Taking these relations into account
\begin{equation}
	W^{(0)}_{JJ} = 1 - \case{1}{2}\Delta^2_2     D
	 -	\Delta_2   \Delta_B   d_J E        
	 -	\case{1}{2}\Delta^2_B [R_1       + d_J R_2],
	\label{eq:betterW0}
\end{equation}
where
\begin{equation}
	\Delta_X = \frac{1}{2m_{Xc}} - \frac{1}{2m_{Xb}} .
	\label{eq:Delta}
\end{equation}
So $W^{(0)}_{JJ}$ is correctly reproduced if $1/m_2$ and $1/m_B$
are adjusted to their continuum values, in particular if the analysis
identifies~$m_2$ with the heavy quark mass.

Another lattice approximant to~$h_+(1)$ and~$h_1(1)$ is given by
double ratios introduced in Refs.~\cite{Hashimoto:1999yp,Simone:1999nv}
\begin{eqnarray}
	R_+ & = &
	\frac{\langle D|V^0_{\text{lat}}|B\rangle%
		  \langle B|V^0_{\text{lat}}|D\rangle}%
		 {\langle D|V^0_{\text{lat}}|D\rangle%
		  \langle B|V^0_{\text{lat}}|B\rangle} ,
	\label{eq:R+def} \\
	R_1 & = &
	\frac{\langle D^*|V^0_{\text{lat}}|B^*\rangle%
		  \langle B^*|V^0_{\text{lat}}|D^*\rangle}%
		 {\langle D^*|V^0_{\text{lat}}|D^*\rangle%
		  \langle B^*|V^0_{\text{lat}}|B^*\rangle} .
	\label{eq:R1def}
\end{eqnarray}
Then~$|h_{+,1}(1)|^2$ are approximated by $\rho_{V_0}^2R_{+,1}$, 
where $\rho_{V_0}^2=Z_{V^{cb}}Z_{V^{bc}}/Z_{V^{cc}}Z_{V^{bb}}$.
To see the advantage of the double ratios, let us rewrite
$W^{(2)}_{JJ}=\bar{W}^{(2)}_{JJ} + \delta W^{(2)}_{JJ}$,
\begin{eqnarray}
	\bar{W}^{(2)}_{JJ} &=&
		\case{1}{2}\Delta^2_3 \left[
			z^{(1,1)}_{V1}\lambda_1 + z^{(1,1)}_{Vs} d_J \lambda_2
		\right], \label{eq:bW2JJ} \\
	\delta W^{(2)}_{JJ} &=& \left(
		\frac{\eta_{VD_\perp^2}^{(2,0)}}{8m_{D_\perp^2c}^2}
	 +	\frac{\eta_{VD_\perp^2}^{(0,2)}}{8m_{D_\perp^2b}^2} -
		\frac{z^{(1,1)}_{V1}}{8m_{3c}^2} - \frac{z^{(1,1)}_{V1}}{8m_{3b}^2}
		\right) \lambda_1
	\nonumber \\
	&+&	\left(
		\frac{\eta_{VsB}^{(2,0)}}{8m_{sB c}^2}
	 +	\frac{\eta_{VsB}^{(0,2)}}{8m_{sB b}^2} -
		\frac{z^{(1,1)}_{Vs}}{8m_{3c}^2} - \frac{z^{(1,1)}_{Vs}}{8m_{3b}^2}
		\right) d_J \lambda_2 . \label{eq:dW2JJ}
\end{eqnarray}
From Eq.~(\ref{eq:DVB}) and the definitions one finds
\begin{equation}
	\rho_{V_0} \sqrt{R_{+,1}} = \eta_V W^{(0)}_{JJ} + \bar{W}^{(2)}_{JJ}
	+ O((\eta_V^{(r,s)}-1)/m^2_Q).
	\label{eq:R+1}
\end{equation}
The contribution of $\delta W^{(2)}_{JJ}$, which stems from the
dimension-five currents, largely cancels.
Hence, the double ratios depend most strongly on $1/m_2$, $1/m_B$, 
and~$1/m_3$, namely the coefficients in~${\cal L}^{(1)}$
and~$V_\mu^{(1)}$.

Equation~(\ref{eq:R+1}) is an important practical result.
If one tolerates errors of order $\alpha_s/m_Q^2$ and~$1/m_Q^3$, then 
$\rho_{V_0}\sqrt{R_{+,1}}$ only requires $m_2=m_B=m_3$ and 
$z^{(1,1)}_V=1$ at the tree level, and details of the 
currents~$V^{(0,2)}$ and~$V^{(2,0)}$ do not matter at all.
With the widely used Sheikholeslami-Wohlert 
action~\cite{Sheikholeslami:1985ij}, this accuracy is easy to 
arrange~\cite{El-Khadra:1997mp,Hashimoto:1999yp}.
In practice an error comes also from~$\eta_V$ and~$\rho_{V_0}$, which 
are available only to two loops~\cite{Czarnecki:1996cf} and one 
loop~\cite{Kronfeld:1998tk}, respectively.
So the recent result~\cite{Hashimoto:1999yp} for $h_{+}(1)$ has a
heavy-quark discretization effect of order $\alpha_s^2$, which could
be reduced by calculating~$\rho_{V_0}$ to two loops.

\subsection{At zero recoil: $h_{A_1}(1)$}

To obtain lattice approximants to~$h_{A_1}(1)$ one must work out
Eq.~(\ref{eq:lat=hqet}) for a $B\to D^*$ transition mediated by the
axial current.
In HQET the currents are as in Eqs.~(\ref{eq:V0})--(\ref{eq:V11}) with 
a factor~$\gamma_5$ inserted in the obvious places.
In this case, the overall factor~$Z_{A^{cb}}$ is conventionally chosen
so that $\delta\eta_{A^{cb}_i}^{\text{lat}}=0$.

A useful matrix element has the $D^*$ spin is aligned along the
$i$~direction and $\bbox{v}=\bbox{v}'=\bbox{0}$.
One finds
\begin{equation}
	\langle D^*|Z_{A^{cb}}A_{\text{lat}}^i|B\rangle =
		\eta_{A^{cb}} W^{(0)}_{01}
		+ \bar{W}^{(2)}_{01} + \delta W^{(2)}_{01}
	\label{eq:DAB}
\end{equation}
in which $\langle D^*_{v}|A^{(0)}|B_{v}\rangle$ yields---after
eliminating $A$, $B$, $C_1$, and $C_3$---
\begin{equation}
	W^{(0)}_{01} = 1
		-	\case{1}{2}\Delta_2(\Delta_2 D   - 2\Theta_B E)        
		-	\case{1}{2}\Delta_B(\Delta_B R_1 -  \Theta_B R_2) 
		-	\frac{1}{2m_{Bc}\,2m_{Bb}} 
			\left(\case{4}{3}R_1 + 2R_2\right)
	\label{eq:W001}
\end{equation}
with
\begin{equation}
	\Theta_X = \frac{1}{2m_{Xc}} + \frac{3}{2m_{Xb}}.
	\label{eq:Theta}
\end{equation}
As before, the zero-recoil matrix element does not depend on the 
dimension-six Lagrangian.
The matrix element $\langle D^*_{v}|A^{(2)}|B_{v}\rangle$ of the 
dimension-five current yields
\begin{eqnarray}
	\bar{W}^{(2)}_{01} & = & \left(
		\case{1}{2}\Delta_3^2 + \case{4}{3}\frac{1}{2m_{3c}2m_{3b}}
		\right) z_{A^{cb}1}^{(1,1)} \lambda_1 -	\left(
	  	\case{1}{2}\Delta_3\Theta_3 - 2\frac{1}{2m_{3c}2m_{3b}} 
	 	\right) z_{A^{cb}s}^{(1,1)} \lambda_2 ,
	\label{eq:W201} \\
	\delta W^{(2)}_{01} & = &
		\left(\frac{\eta_{A^{cb}D_\perp^2}^{(2,0)}}{8m_{D_\perp^2c}^2}
		  +   \frac{\eta_{A^{cb}D_\perp^2}^{(0,2)}}{8m_{D_\perp^2b}^2}
		  -   \frac{z_{A^{cb}1}^{(1,1)}}{8m_{3c}^2}
		  -   \frac{z_{A^{cb}1}^{(1,1)}}{8m_{3b}^2} \right) \lambda_1 
	\label{eq:dW201} \\
	&-&	\left(\frac{\eta_{A^{cb}sB}^{(2,0)}}{8m_{sB c}^2}
		  -   \frac{3\eta_{A^{cb}sB}^{(0,2)}}{8m_{sB b}^2}
		  -   \frac{z_{A^{cb}s}^{(1,1)}}{8m_{3c}^2}
		  +   \frac{3z_{A^{cb}s}^{(1,1)}}{8m_{3b}^2} \right) \lambda_2 ,
	\nonumber
\end{eqnarray}
after grouping terms as in Eqs.~(\ref{eq:bW2JJ}) and~(\ref{eq:dW2JJ}).

Reference~\cite{Simone:1999nv} introduces a third double ratio
\begin{equation}
	R_{A_1} = 
	\frac{\langle D^*|A^i_{\text{lat}}|B\rangle%
		  \langle B^*|A^i_{\text{lat}}|D\rangle}%
		 {\langle D^*|A^i_{\text{lat}}|D\rangle%
		  \langle B^*|A^i_{\text{lat}}|B\rangle}.
	\label{eq:RA1def}
\end{equation}
After substituting for each matrix element the foregoing expressions one
finds
\begin{equation}
	\rho_{A} \sqrt{R_{A_1}} = \check{\eta}_{A^{cb}} \check{W}^{(0)}_{01}
		+ \check{W}^{(2)}_{01} + O((\check{\eta}_A^{(r,s)}-1)/m^2_Q), 
	\label{eq:RA1} \\
\end{equation}
where $\rho_A^2=Z_{A^{cb}}Z_{A^{bc}}/Z_{A^{cc}}Z_{A^{bb}}$, 
$\check{\eta}_{A^{cb}}^2=%
\eta_{A^{cb}}\eta_{A^{bc}}/\eta_{A^{cc}}\eta_{A^{bb}}$, and
\begin{eqnarray}
	\check{W}^{(0)}_{01} & = & 1
		- \case{1}{2}  \Delta^2_2     D
		-	\Delta_2     \Delta_B       E        
		+	\case{1}{6}  \Delta^2_B (R_1 + 3 R_2)
	\label{eq:cW001} \\
	\check{W}^{(2)}_{01} & = & - \case{1}{6} \Delta^2_3 
		(  \check{z}_{A^{cb}1}^{(1,1)}\lambda_1
		+ 3\check{z}_{A^{cb}s}^{(1,1)}\lambda_2),
	\nonumber
\end{eqnarray}
where
$\check{z}_{A^{cb}}=\check{\eta}_{A^{cb}}z_{A^{cb}}/\eta_{A^{cb}}$.
As before the contribution~$\delta W^{(2)}_{01}$ of the dimension-five
currents largely cancels and the double ratio depends most strongly
on~$1/m_2$, $1/m_B$, and~$1/m_3$, namely the coefficients
in~${\cal L}^{(1)}$ and~$A_\mu^{(1)}$.

Note, however, that $\rho_A\sqrt{R_{A_1}}$ does not 
yield~$h_{A_1}(1)$ but
\begin{equation}
	\rho_A^2R_{A_1} = \frac{%
		h_{A_1}^{B\to D^*}(1)h_{A_1}^{D\to B^*}(1)}{%
		h_{A_1}^{D\to D^*}(1)h_{A_1}^{B\to B^*}(1)} .
\end{equation}
Nevertheless, if the action and currents are tuned so that $1/m_2$,
$1/m_B$, $1/m_3$, and~$z^{(1,1)}_j$ match the usual HQET (to a desired
accuracy), the three double ratios $R_+$, $R_1$, and $R_{A_1}$ can
be combined to yield the $1/m_Q^2$ contribution to~$h_{A_1}(1)$.
For example, if one tolerates errors of order~$\alpha_s/m_Q^2$, as
well as~$1/m_Q^3$, one only requires $m_2=m_B=m_3$ at the tree level,
and one may set $z_V^{(1,1)}=\check{z}_A^{(1,1)}=1$.
Then, dropping the distinction between $m_2$, $m_B$, and $m_3$, the 
double ratios are
\begin{eqnarray}
	\bar{\rho}_V^2 R_+ & = & 1 - 2\Delta^2 \ell_P ,
		\label{eq:ellP} \\
	\bar{\rho}_V^2 R_1 & = & 1 - 2\Delta^2 \ell_V ,
		\label{eq:ellV} \\
	\bar{\rho}_A^2 R_{A_1} & = & 1 - 
		\Delta^2(\ell_P+\ell_V+\ell_A) , \label{eq:ellPVA}
\end{eqnarray}
where $\bar{\rho}_V=\rho_V/\eta_V$, 
$\bar{\rho}_A=\rho_A/\check{\eta}_A$, and 
\begin{eqnarray}
	2\ell_P & = & D + R_1 - \lambda_1 + 3(2E + R_2 - \lambda_2)
	\label{eq:ellPmx} \\
	2\ell_V & = & D + R_1 - \lambda_1 -  (2E + R_2 - \lambda_2)
	\label{eq:ellVmx} \\
	\ell_A & = & \case{4}{3}(\lambda_1 - R_1) + 2(\lambda_2 - R_2).
	\label{eq:ellAmx}
\end{eqnarray}
In the approximation being considered the desired form factor is
\begin{equation}
	\frac{h_{A_1}(1)}{\eta_A} = 1 - 
		\Delta \left(\frac{\ell_V}{2m_c} - \frac{\ell_P}{2m_b}\right) +
		\frac{\ell_A}{2m_c\,2m_b}
	\label{eq:DABell}.
\end{equation}
By fitting Eqs.~(\ref{eq:ellP})--(\ref{eq:ellPVA}) one can 
extract~$\ell_P$, $\ell_V$, and $\ell_P+\ell_V+\ell_A$, and then one 
has the information necessary to reconstitute~$h_{A_1}(1)$.
As with~$h_+(1)$ there are, in practice, further errors
because~$\eta_A$ and~$\rho_A$ are available only at finite-loop
order~\cite{Czarnecki:1996cf,Kronfeld:1998tk}.

\subsection{Near zero recoil: $h_-(1)$}
\label{sec:near}

To extract~$h_-(1)$ matrix elements with non-zero velocity transfer
are needed, and some new features appear in the analysis.
For example, lattice approximants to $h_-$ receive a contribution from
the term in~$V^{(0)}$ proportional to~$\beta_{V_\mu}^{\text{lat}}$.
Similar ``$\beta$-like'' terms omitted from
Eqs.~(\ref{eq:V01})--(\ref{eq:V11}) also make contributions.
We shall not write out all these terms but indicate instead how they 
contribute to matrix elements.

Suppose one extracts the form factor from a matrix element with
$\bbox{v}'=-\bbox{v}$, pointing in the $i$ direction.
Then
\begin{equation}
	\langle D(-\bbox{v})|V_i|B(\bbox{v})\rangle = v_i 
		\left\{\beta_{V_i}^{\text{lat}} W^{(0)}_{00} +
		Y^{(2)}_{00} + X^{(1)}_{00} + X^{(2)}_{00} \right\},
	\label{eq:DViB}
\end{equation}
where $W^{(0)}_{00}$ is given in Eq.~(\ref{eq:betterW0}), 
and   $Y^{(2)}_{00}$ is like $W^{(2)}_{00}$ but with $\beta$-like 
coefficients replacing $\eta_V^{(2,0)}$, $\eta_V^{(0,2)}$,
and~$z_V^{(1,1)}$.
The expression in braces is a lattice approximant to~$h_-(w)$.
For infinitesimal~$v_i$ the matrix element
$\langle D_{v'}|V^{(1)}|B_v\rangle$ yields
\begin{eqnarray}
	X^{(1)}_{00} & = &
		\left(\frac{\eta_V^{(1,0)}}{2m_{3c}}
		  -   \frac{\eta_V^{(0,1)}}{2m_{3b}} \right)
		\left[2\xi_3(1) - \bar{\Lambda}
		  +  2\Sigma_2 \Xi_3(1) + 2\Sigma_B \Xi_-(1)\right] 
	\nonumber \\
	&-&	\left(\frac{\eta_V^{(1,0)}}{2m_{3c}}
		  +   \frac{\eta_V^{(0,1)}}{2m_{3b}} \right)
		\left[\Delta_2 \tilde{\phi}_0(1) +
			  \Delta_B \tilde{\phi}_-(1)\right],
	\label{eq:X100}
\end{eqnarray}
with $\Delta_2$, $\Delta_B$ as in Eq.~(\ref{eq:Delta}) and
\begin{equation}
	\Sigma_X = \frac{1}{2m_{Xc}} + \frac{1}{2m_{Xb}}.
	\label{eq:Sigma}
\end{equation}
The chromoelectric part of $\langle D_{v'}|V^{(2)}|B_v\rangle$ yields
\begin{equation}
	X^{(2)}_{00} = \left(
		\frac{\eta_{V\alpha E}^{(2,0)}}{4m_{\alpha Ec}^2} -
		\frac{\eta_{V\alpha E}^{(0,2)}}{4m_{\alpha Eb}^2}\right)
			\frac{2}{3}\left[\lambda_1 + 3\lambda_2\right] .
	\label{eq:X200}
\end{equation}
The coefficient factors, together with Table~\ref{tab:not}, make
clear the origin of each term.
The infinite-mass matrix elements $\bar\Lambda$, $\lambda_1$,
and $\lambda_2$ are exactly those introduced earlier, and the new
ones $\xi_3(1)$, $\Xi_3(1)$, $\Xi_-(1)$, $\tilde\phi_0(1)$, and
$\tilde\phi_-(1)$ are introduced in Appendix~\ref{app:near0}.
As before, the dimension-six effective Lagrangian drops out,
but the dimension-five currents contribute in several places:
in~$W^{(0)}_{00}$, $Y^{(2)}_{00}$, and~$X^{(2)}_{00}$.
Further operators $-i\bar{c}_{v'}D^\mu_\perp b_v$,
$i\bar{c}_{v'}\loarrow{D}^\mu_{\perp'}b_v$, $\bar{c}_{v'}E^\mu b_v$,
and $\bar{c}_{v'}E'^\mu b_v$, whose coefficients vanish at the tree
level, modify the short-distance coefficients of~$\bar{\Lambda}$
in~$X^{(1)}_{00}$ and of~$\lambda_1$ in~$X^{(2)}_{00}$.
Thus, many short-distance coefficients influence the accuracy of
Eq.~(\ref{eq:DViB}).

The main drawback of Eq.~(\ref{eq:DViB}) is, however, the requirement 
$\bbox{v}'=-\bbox{v}$ for hadrons of unequal mass.
Numerical calculations employ a finite volume and, hence, discrete 
momentum.
Moreover, with the many ``masses'' the relation between
momentum and velocity is not plain.
To remove these ambiguities Ref.~\cite{Hashimoto:1999yp} introduced
another double ratio
\begin{equation}
	R_- =
	\frac{\langle D|V^i_{\text{lat}}|B\rangle}%
		 {\langle D|V^0_{\text{lat}}|B\rangle}
	\frac{\langle D|V^0_{\text{lat}}|D\rangle}%
		 {\langle D|V^i_{\text{lat}}|D\rangle} .
	\label{eq:R-def}
\end{equation}
In the spatial matrix elements, the initial state is at rest and the
final state has a small velocity in the $i$ direction; in the temporal
matrix elements, initial and final states both are at rest.
In continuum QCD, the analogous first ratio is
[\emph{cf.}\ Eq.~(\ref{eq:h+h-})]
\begin{equation}
	\frac{\langle D|{\cal V}^i|B\rangle}%
		 {\langle D|{\cal V}^0|B\rangle} = \case{1}{2} v'_i
		 \left[1-\frac{h_-(1)}{h_+(1)}\right],
	\label{eq:R-num}
\end{equation}
to first order in~$v'_i$, and the second is
\begin{equation}
	\frac{\langle D|{\cal V}^i|D\rangle}%
		 {\langle D|{\cal V}^0|D\rangle} = \case{1}{2} v'_i,
	\label{eq:R-den}
\end{equation}
because in the elastic case $h_+(1)=1$ and~$h_-(w)=0$.
Thus, with a suitable adjustment of the lattice currents, one can 
use~$R_-$ to obtain a lattice approximant to~$h_-(1)/h_+(1)$.

In the double ratio of lattice currents the (mass-dependent)
factors~$Z_V$ cancel.
With the results of Appendix~\ref{app:near0}, and noting that
$\eta_{V^{cc}}=1$ and $\beta_{V_i^{cc}}^{\text{lat}}=0$,
\begin{equation}
	R_- = \frac{\eta_{V^{cb}} + \delta\eta_{V^{cb}_i}^{\text{lat}}
		- [ \beta_{V_i^{cb}}^{\text{lat}} + Y^{(2)}_{00}
		+ X^{(1)}_{00} + X^{(2)}_{00} ]
		+ \delta\bar\eta_{V^{cb}}^{(2)} W^{(2)}_{00}}{%
		\eta_{V^{cb}} (1 + \delta\eta_{V^{cc}_i}^{\text{lat}}
		+ \delta\bar\eta_{V^{cc}}^{(2)} W^{(2)}_{00})} 
		+ O(1/m_Q^3) ,
	\label{eq:R-exp}
\end{equation}
where $\delta\bar\eta_{V^{cb}}^{(2)}$ is a combination
of~$\delta\eta_{V_i}^{(r,s)}/\eta_V^{(r,s)}$ and
$(\delta\eta_{V_i}^{\text{lat}}-\beta_{V_i}^{\text{lat}})/\eta_V$.
Here~$W^{(2)}_{00}$, $X^{(s)}_{00}$, and~$Y^{(2)}_{00}$ are precisely
as above, though in the denominator $W^{(2)}_{00}$ is evaluated with
flavor~$c$ in both final and initial states.
As with the other double ratios, one would like to extract the
long-distance information from~$R_-$.
To do so one must have a way to calculate
the short-distance coefficients, either to
adjust them so~$\beta_{V_i}^{\text{lat}}=\beta_V$
and~$\delta\eta_{V_i}^{\text{lat}}=\delta\bar\eta_V^{(2)}=0$, or to
constrain a fit.

A simple version of the latter strategy is available if one tolerates
errors in~$h_-(1)$ of order $\alpha_s/m_Q$ and~$\alpha_s/m_Q^2$.
Then it is enough to adjust $m_3=m_2=m_B$ at the tree level,
one may set~$\eta_V^{(r,0)}=\eta_V^{(0,s)}=1$, and one may
neglect~$Y^{(2)}_{00}$ and~$\delta\bar\eta_V^{(2)}W^{(2)}_{00}$.
In the approximation at hand, Eq.~(\ref{eq:R-exp}) can be rearranged
to yield
\begin{equation}
	\eta_{V^{cb}}[1 - (1+\delta\eta_{V^{cc}_i}^{\text{lat}})R_-]
		+ \delta\eta_{V_i^{cb}}^{\text{lat}}
		- \beta_{V_i^{cb}}^{\text{lat}} =
	X^{(1)}_{00} + X^{(2)}_{00} .
	\label{eq:bX}
\end{equation}
Setting~$m_3=m_2=m_B$, but keeping $m_{\alpha E}$ distinct, 
Eqs.~(\ref{eq:X100}) and.~(\ref{eq:X200}) yield
\begin{equation}
	X^{(1)}_{00} + X^{(2)}_{00} = \Delta \ell_-^{(1)}
		+ \Delta\Sigma \ell_-^{(2)}
		+ \case{2}{3} \Delta_{\alpha E}\Sigma_{\alpha E}
			[\lambda_1+3\lambda_2],
	\label{eq:XX}
\end{equation}
where
\begin{eqnarray}
	\ell_-^{(1)} & = & 2\xi_3(1) - \bar\Lambda, \\
	\ell_-^{(2)} & = & 2\Xi_3(1) + 2\Xi_-(1) 
		- \tilde{\phi}_0(1) - \tilde{\phi}_-(1) .
\end{eqnarray}
One may fit the left-hand side of Eq.~(\ref{eq:bX}) to the right-hand
side of Eq.~(\ref{eq:XX}) with $1/m^2_{\alpha E}$ at the easily
obtained tree level.
After the fit one may reconstitute~$h_-(1)$ from
\begin{equation}
	\frac{h_-(1)}{h_+(1)} = \beta_V + 
	 \Delta \ell_-^{(1)}
		+ \Delta \Sigma\left[\ell_-^{(2)} + \case{2}{3}
			\left(\lambda_1+3\lambda_2\right)
		\right].
\end{equation}
In practice, there are also errors of order~$\alpha_s^n$ because the 
coefficients~$\eta_V$, $\beta_V$, $\delta\eta_V^{\text{lat}}$, 
and~$\beta_V^{\text{lat}}$ are available only to a finite loop order.
Note that the matrix element $\lambda_1+3\lambda_2$ appears also as
the $1/m_Q$ correction to the pseudoscalar meson mass, \emph{cf.}\
Eq.~(\ref{eq:M1}), so a simultaneous fit may turn out to be useful.

\section{Leptonic Decays}
\label{sec:fB}

A straightforward application of the trace formalism gives the
first-order heavy-quark expansion of the matrix element in leptonic
decays.
The result for lattice QCD is in Ref.~\cite{Kronfeld:1995nu}, but
for completeness the derivation is given here.

With the states normalized as in Eq.~(\ref{eq:normstates}), the QCD 
amplitudes appearing in leptonic decays of heavy-light pseudoscalar
and vector mesons can be written
\begin{eqnarray}
	\langle 0|{\cal A}^\mu| H(v)\rangle & = & 
		i v^\mu      \phi_H/\sqrt{2}, \label{eq:phiP} \\
	\langle 0|{\cal V}^\mu| H^*(v,\epsilon)\rangle & = & 
		\epsilon^\mu \phi_{H^*}/\sqrt{2}, \label{eq:phiV}
\end{eqnarray}
where ${\cal V}^\mu$ and ${\cal A}^\mu$ are now the vector and axial 
vector currents with a light and a heavy quark, and $H$ ($H^*$) is
the pseudoscalar (vector) meson with heavy flavor~$h$.
The relation between the parameter~$\phi_H$ and the conventional
pseudoscalar meson decay constant is 
\begin{equation}
	\phi_H = f_H\sqrt{M_H}.
	\label{eq:fP}
\end{equation}
There are several conventions for defining the vector meson decay
constant, but only $\phi_{H^*}$ is considered here.

In lattice gauge theory the decay constants are approximated with
matrix elements of lattice currents $Z_{V^{qh}}V^{qh}$ 
and~$Z_{A^{qh}}A^{qh}$ with the same quantum numbers as~${\cal V}^\mu$ 
and~${\cal A}^\mu$.
As before, they are not made explicit, to allow for a variety of
choices.
The underlying currents are described by HQET currents,
\begin{eqnarray}
	Z_{V^{qh}} V^{qh}_\mu & \doteq & 
		\eta_{V^{qh}} \bar{q} i\gamma_\mu h_v 
	 +	\zeta_{V^{qh}} v_\mu \bar{q}h_v 
	 -	\frac{\eta_{V^{qh}}^{(0,1)}}{2m_3}\bar{q}i\gamma_\mu 
	 	{\kern+0.1em /\kern-0.65em D}_\perp h_v + \cdots
	\label{eq:hlV} \\
	Z_{A^{qh}} A^{qh}_\mu & \doteq & 
		\eta_{A^{qh}}  \bar{q} i\gamma_\mu\gamma_5 h_v 
	 +	\zeta_{A^{qh}} v^\nu \bar{q} i\sigma_{\mu\nu}\gamma_5 h_v 
	 -	\frac{\eta_{A^{qh}}^{(0,1)}}{2m_3}\bar{q}i\gamma_\mu \gamma_5 
		{\kern+0.1em /\kern-0.65em D}_\perp h_v + \cdots
\label{eq:hlA}
\end{eqnarray}
where~$\bar q$ is a light anti-quark field.
The coefficient~$1/2m_3$ is defined through the degenerate-mass
heavy-heavy vector current, and $\eta_j^{(0,1)}$ captures the remaining
radiative corrections.
At the tree level $\eta_j^{(0,1)}=1$.
The coefficients~$\zeta_j$ vanish at the tree level, and the operators 
that they multiply do not affect~$\phi_{H^{(*)}}$.
Additional dimension-four operators, whose coefficients vanish at the
tree level, are not written~out.

The static limit is given by the matrix element of the first term of 
the HQET currents:
\begin{eqnarray}
	\langle 0|\bar{q}i\Gamma_\mu h_v|h_vJ\rangle &=&
		- \case{1}{2} \phi_\infty
			\tr\left[i\Gamma_\mu{\cal M}_J\right] \nonumber \\
	& = & i\omega_\mu \phi_\infty/\sqrt{2},
	\label{eq:phi0}
\end{eqnarray}
where $\Gamma_\mu=\gamma_\mu\gamma_5$ or~$\gamma_\mu$ and
$\omega_\mu=v_\mu$ or~$-i\epsilon_\mu$, for $J=0$ or~1.
The constant~$\phi_\infty/2$ is introduced to parametrize the light 
degrees of freedom;
in the static limit, $\phi_H=\phi_{H^*}=\phi_\infty$.
As with the quantities introduced in Secs.~\ref{sec:masses} 
and~\ref{sec:semi}, $\phi_\infty$ differs from its continuum limit, 
but the difference stems only from the light degrees of freedom.

At order $1/m_Q$ there are three contributions to~$\phi_{H^{(*)}}$, from the 
kinetic and chromomagnetic energy, and from the correction to the 
current.
They take the form
\begin{eqnarray}
	\langle 0|Z_j j_\mu|H^{(*)}\rangle & = &
		\eta_j \langle 0|\bar{q}i\Gamma_\mu h_v|h_vJ\rangle 
		 +	\frac{\eta_j}{2m_2} \int d^4x\,
		 \langle 0|T\, \bar{q}i\Gamma_\mu h_v(0)
		 	{\cal O}_2(x)|h_vJ\rangle^\star
	\label{eq:J1/M} \\
		&+&	\frac{\eta_j}{2m_B} \int d^4x\,
		\langle 0|T\, \bar{q}i\Gamma_\mu h_v(0)
			{\cal O}_B(x)|h_vJ\rangle^\star
		 -	\frac{\eta_j^{(0,1)}}{2m_3}\langle 0|\bar{q}i\Gamma_\mu
			{\kern+0.1em /\kern-0.65em D}_\perp h_v|h_vJ\rangle. 
	\nonumber
\end{eqnarray}
Spin-dependent factors may be obtained with the trace formalism.
One has
\begin{eqnarray}
	\int d^4x\,
	\langle 0|T \bar{q}i\Gamma_\mu h_v\,{\cal O}_2(x)|h_vJ\rangle &=& 
		- \case{1}{2} (- \phi_\infty A_2) 
			\tr\left[i\Gamma_\mu{\cal M}_J\right] \nonumber \\
	&=& - \frac{   \phi_\infty A_2}{\sqrt{2}} i\omega_\mu
	\label{eq:phi2}
\end{eqnarray}
\begin{eqnarray}
	\int d^4x\,
	\langle 0|T \bar{q}i\Gamma_\mu h_v\,{\cal O}_B(x)|h_vJ\rangle &=&
		- \case{1}{12}(- \phi_\infty A_B) 
		\tr\left[i\Gamma_\mu\sigma^{\rho\sigma}{\cal M}_J
		\sigma_{\alpha\beta}\right] \eta^\alpha_\rho \eta^\beta_\sigma
		\nonumber \\
	&=& - \frac{d_J\phi_\infty A_B}{3\sqrt{2}} i\omega_\mu
	\label{eq:phiB}
\end{eqnarray}
\begin{eqnarray}
	\langle 0|\bar{q}i\Gamma_\mu
	{\kern+0.1em /\kern-0.65em D}_\perp h_v|h_vJ\rangle &=& 
		- \case{1}{2}\phi_\infty A_3
	\tr\left[i\Gamma_\mu\gamma_\perp^\alpha{\cal M}_J\gamma_\alpha\right] 
	\nonumber \\
	&=& + \frac{d_J \phi_\infty A_3}{\sqrt{2}} i\omega_\mu
	\label{eq:phi3}
\end{eqnarray}
where $A_2$, $A_B$, and $A_3$ parametrize the light degrees of 
freedom, and $d_H=3$, $d_{H^*}=-1$.
Combining the equations of motion and heavy-quark symmetry,
\begin{equation}
	A_3=\case{1}{3} (\bar{\Lambda}-m_q), 
	\label{eq:A3}
\end{equation}
where~$m_q$ is the mass of the light quark~\cite{Falk:1992fm}.
Combining Eqs.~(\ref{eq:phi0})--(\ref{eq:A3})
\begin{equation}
	\phi_{H^{(*)}} = \phi_\infty\left[\eta_j\left(1 -
		\frac{A_2}{2m_2} - \frac{d_J}{3}\frac{A_B}{2m_B}\right) -
		\eta_j^{(0,1)}\frac{d_J}{3}
		\frac{\bar{\Lambda}-m_q}{2m_3}\right].
	\label{eq:phi}
\end{equation}
As expected on the general grounds outlined in 
Sec.~\ref{sec:properties}, the rest mass does not appear.
Previously this had been shown only by explicit
calculation~\cite{Falk:1992fm}.
Like the mass formula~(\ref{eq:M1}), this result is simple enough that
it could have been written down upon inspection of the corresponding
continuum formula~\cite{Neubert:1992fk,Falk:1992fm}.

To obtain the correct static limit of the decay constants, one must 
adjust the normalization factors~$Z_j$ to yield~$\eta_j$ in the 
leading terms.
This is known at the one-loop level for the
Wilson~\cite{Kuramashi:1998tt} and Sheikholeslami-Wohlert
actions~\cite{Ishikawa:1997xh}.
Similarly, to obtain the $1/m_Q$ corrections, one must adjust the 
lattice action and currents so that $m_2=m_B=m_3=m_Q$, which is easy 
at the tree level.
With these choices, Eq.~(\ref{eq:phi}) predicts that the heavy-light
decay constants should depend mildly on the lattice spacing.
Explicit calculation supports this
prediction~\cite{Aoki:1998ji,El-Khadra:1998hq}.
On the other hand, when not all these choices are made, the dependence
on the lattice spacing could be more pronounced, because then~$1/m_3$
or~$1/m_B$ could vary rapidly with~$m_Qa$.
Explicit calculation supports this prediction
too~\cite{Bernard:1998xi}.

\section{Discussion and Conclusions}
\label{sec:conclusions}

Two themes run through Symanzik's application of effective field
theory to the study of cutoff effects.
The first is descriptive~\cite{Symanzik:1979ph}.
The local effective Lagrangian organizes deviations from the continuum 
limit through a series of higher-dimension operators, multiplied with 
certain coefficients.
When the higher-dimension terms are small, they can be treated as 
perturbations, and their influence can be propagated from the effective 
Lagrangian to physical quantities.
The second theme turns the description into a 
weapon~\cite{Symanzik:1983dc}.
Details of the underlying lattice action alter the effective 
Lagrangian only via the short-distance coefficients.
If a given action leads to a reduced (or vanishing) coefficient, then 
the process independence of the coefficient guarantees that its 
associated operator has a reduced (or vanishing) effect on all 
observables.

The two themes also run through the application of HQET to lattice QCD.
The concrete results---the expansions given in Eq.~(\ref{eq:M1}),
Eqs.~(\ref{eq:DVB})--(\ref{eq:dW2JJ}),
Eqs.~(\ref{eq:DAB})--(\ref{eq:dW201}),
Eqs.~(\ref{eq:DViB})--(\ref{eq:X200}), and
Eq.~(\ref{eq:phi})---describe the deviations from the static limit of
the mass, semileptonic form factors, and decay constant of heavy-light
mesons.
These descriptions hold, as always in HQET, when momentum transfers 
are much smaller than the heavy quark mass(es).
Details of the lattice alter the validity of the description
superficially: they merely change the short-distance coefficients.
On the other hand, the details alter the utility of the description
greatly: if a coefficient is tuned correctly, to some accuracy, in
one observable, then its associated operator contributes correctly,
to that accuracy, in all observables.
In all examples, one sees that the leading $1/m_Q$ dependence is 
reproduced correctly if the short-distance coefficients $1/m_2$, 
$1/m_B$, and $1/m_3$ are adjusted correctly.
These conditions can be obtained, respectively, through suitable
adjustments of the bare mass, of the ``clover'' coupling in the
Sheikholeslami-Wohlert action, and of a tunable parameter in the
current.

It may be worthwhile to contrast the formalism developed here with
other methods for treating heavy quarks in lattice gauge theory.
One approach is to derive HQET or NRQCD in the continuum and discretize
the result.
In fact, both effective theories were originally formulated with this 
idea in
mind~\cite{Caswell:1986ui,Lepage:1987gg,Lepage:1992tx,Eichten:1990zv}.
The resulting lattice theory has ultraviolet divergences that are more
severe than those of~QCD, so one must either keep $a^{-1}\sim m_Q$ and 
employ a highly improved lattice 
action~\cite{Lepage:1987gg,Lepage:1992tx} or restrict one's attention 
to the leading term of the infinite-mass limit~\cite{Maiani:1992az}.
The approach developed here and in Ref.~\cite{El-Khadra:1997mp}
examines the large-mass limit of Wilson fermions, and as $a\to 0$ the 
only ultraviolet divergences that are encountered are those of~QCD.

Another approach is based on lattice actions that are asymmetric under
interchange of the temporal and spatial axes~\cite{El-Khadra:1997mp}.
With a suitable adjustment of the asymmetric couplings, the physics 
can be made relativistically covariant.
For example, one can adjust the action so that $m_1=m_2$.
Cutoff effects can be analyzed either with Symanzik's effective
Lagrangian, provided one retains the full dependence on~$m_Qa$ in
the coefficient functions, or with the HQET description developed here.
Initial results~\cite{Sroczynski:1999he} with the asymmetric action
indicate that the Symanzik and HQET interpretations give the same
physical results.

Many papers have followed an \emph{ad hoc} combination of Symanzik
and heavy-quark effective theories.
Numerical data are generated with artificially small heavy-quark
masses, to reduce~$m_Qa$.
Then these data are extrapolated up in mass guided by the (continuum)
$1/m_Q$ expansion.
In practice, however, it is hard to find a region with $m_Qa\ll 1$,
for Symanzik's analysis genuinely to apply to cutoff effects, and
$\Lambda_{\text{QCD}}/m_Q\ll 1$, for HQET genuinely to apply to the
mass dependence.
Often neither asymptotic condition realistically describes the
numerical data.
The description developed in this paper naturally applies to the 
subset of such data where HQET is indeed valid, so these data could be 
reanalyzed in light of the expansions given above.

One might also imagine reducing the lattice spacing~$a$ by an order
of magnitude or so.
In this regime, the pictures painted by HQET and Symanzik's effective 
Lagrangian become indistinguishable from each other, even for the 
bottom quark~\cite{El-Khadra:1997mp}.
The brute-force approach is costly, however.
Processor requirements grow as~$a^{-5}$ (if not faster) and memory 
as~$a^{-4}$.
For $B$ physics it makes more sense to invest steady improvements in
computers into removing the quenched approximation, rather than into
a radical reduction of the lattice spacing.

Some readers may consider the proliferation of short-distance 
coefficients for higher-dimension operators to be impractical.
The proliferation is genuine: it appears to the same extent in lattice 
NRQCD and to almost the same extent in the usual~HQET.
In many cases, however, the uncertainty from one-loop coefficients is 
smaller than other numerical 
uncertainties~\cite{Hashimoto:1999yp,Simone:1999nv}.
Moreover, with limited computer resources the approach of this paper 
is more practical than a brute-force reduction of the lattice spacing 
and on sounder theoretical footing than \emph{ad hoc} fitting 
procedures.

A gap left by this paper is the calculation of the short-distance
coefficients, which depend on the lattice action.
Detailed calculations have been and will be addressed 
elsewhere~\cite{Mertens:1998wx,Kronfeld:1998tk}.
The coefficients can be obtained with some accuracy via perturbation 
theory in the gauge coupling.
There are, for example, general formulae, valid to every order in 
perturbation theory, relating the self energy of the underlying 
lattice theory to the first two coefficients of the effective 
Lagrangian, $m_1$ and $1/m_2$~\cite{Mertens:1998wx}.
Similarly, radiative corrections to the currents are related to the
(on-shell) vertex function~\cite{Kronfeld:1998tk}.
Beyond the one-loop level the calculations will not be easy,
but at least they are well defined.

An even better strategy would be to devise non-perturbative methods for
tuning, if not explicitly calculating, the short-distance physics.
For example, heavy-quark expansions of a hadron's kinetic mass,
chromomagnetic mass, \emph{etc.}, would be useful, because
with them one could remove HQET scheme dependence.
Other possibilities might mimic strategies invented for light quarks,
such as imposing---at finite lattice spacing---identities of the
continuum limit.
For heavy quarks, reparametrization invariance~\cite{Luke:1992cs},
which is closely related to Lorentz invariance and heavy-quark
symmetry, may be helpful.

The heavy-quark expansions in this paper are just the beginning.
A~wide variety of physically interesting observables have been studied
with the usual HQET, and matrix elements of the infinite-mass limit are
almost always needed.
One can re-analyze each observable with the modified coefficients
appropriate to the HQET description of lattice gauge theory, to find
out how a direct lattice calculation compares to the continuum.
Furthermore, it might be possible to extract parameters such
as~$\bar\Lambda$ and~$\lambda_1$ by calculating the short-distance
coefficients (in a suitable scheme) and fitting lattice data.
The idea is similar to a proposal~\cite{Dawson:1998ic} for extracting
kaon matrix elements from current-current correlation functions
$\langle J(x)J(0)\rangle$.
(A~significant difference is that here the ratio~$m_Qa$ of short
distances is treated exactly, whereas in Ref.~\cite{Dawson:1998ic}
the analogous ratio~$a/x$ is presumed small.)
Determinations of~$\bar\Lambda$ and~$\lambda_1$ are intriguing,
because they also appear in heavy-quark expansions of inclusive
processes~\cite{Falk:1994dh,Bigi:1995ga,Uraltsev:1998bk}.


\acknowledgments
I~thank Shoji Hashimoto, Aida El-Khadra, and Paul Mackenzie for
collaboration on related work~\cite{El-Khadra:1997mp,Kronfeld:1998tk},
Laurent Lellouch for probing questions at Lattice '99,
and Zoltan Ligeti and Dan Pirjol for helpful remarks.
Fermilab is operated by Universities Research Association Inc.,
under contract with the U.S. Department of Energy.

\appendix
\section{Traces for Semi-leptonic form factors}
\label{app:traces}

This appendix gives the traces needed to express the semi-leptonic form 
factors, at zero recoil.
Matrix elements with $v'=v$ are considered first, 
in Appendix~\ref{app:at0}.
They enter into~$h_+(1)$, $h_1(1)$, and~$h_{A_1}(1)$.
To extract $h_-(1)$ one must take~$v'$ different from~$v$, focus on 
terms multiplying $\case{1}{2}(v'-v)^\mu$, and then set~$w=1$;
\emph{cf.}\ Appendix~\ref{app:near0}.

\subsection{At zero recoil}
\label{app:at0}
The traces needed to express matrix elements used to obtain~$h_+(1)$, 
$h_1(1)$, and~$h_{A_1}(1)$ are worked out here.
One finds no contribution of the types $\langle j^{(1)}\rangle$ and
$\langle j^{(1)}{\cal L}^{(1)}\rangle$ when $w=1$.

\subsubsection{Contributions from $\langle j^{(0)}\rangle$}
At leading order in the heavy-quark expansion, all matrix elements are 
written
\begin{equation}
	\langle c_{v'}J'|\bar{c}_{v'}\Gamma b_v|b_vJ\rangle = 
		-\tr\{\bar{\cal M}_{J'} \Gamma {\cal M}_J\} \xi(w)
	\label{eq:xi}
\end{equation}
where $w=-v'\cdot v$.
The spin dependence factors completely; there is only one
function~$\xi(w)$ to parametrize the light degrees of freedom.
At zero recoil the current~$iv^\mu\bar{c}_vb_v$ is the 
Noether current of heavy-quark flavor symmetry.
The associated charge changes nothing but the heavy-quark flavor,
namely
\begin{equation}
	\int d^3y \langle c_vJ| iv^0\bar{c}_vb_v(y) = \langle b_vJ|,
	\label{eq:MQ=M}
\end{equation}
and hence $\xi(1)=1$.
Fortunately, this conclusion does \emph{not} depend on the conservation
of the current in the underlying theory, because for lattice QCD one
usually computes the transition with a current that is not conserved.
(That is why~$Z_V$ is written explicitly.)
The violation of current conservation is a short-distance effect,
however, so it can appear only in the short-distance coefficients.

The matrix elements of interest are
\begin{eqnarray}
	\langle c_{v'}0|\bar{c}_{v'} i\gamma^\mu b_v|b_v0\rangle & = &
		\case{1}{2}(v'+v)^\mu \xi(w),
	\label{eq:PVP} \\
	\langle c_{v'}1|\bar{c}_{v'}i\gamma^\mu
		b_v|b_v1\rangle & = &
		\case{1}{2}(v'+v)^\mu \overline{\epsilon'}\!\cdot\!\epsilon\, \xi(w),
	\label{eq:VVV} \\
	\langle c_{v'}1|\bar{c}_{v'}i\gamma^\mu\gamma_5 b_v
		|b_v0\rangle & = &
		\case{1}{2}[(1+w)i\overline{\epsilon'}^\mu +
		i\overline{\epsilon'}\cdot v v^{\prime\mu} ]\xi(w).
	\label{eq:VAP} \\
	\langle c_{v'}0|\bar{c}_{v'} i\gamma^\mu\gamma_5 b_v
		|b_v1\rangle & = &
		-\case{1}{2}[(1+w)i\epsilon^\mu +
		i\epsilon\!\cdot\!v'\,v^\mu ]\xi(w).
	\label{eq:PAV}
\end{eqnarray}
In Eqs.~(\ref{eq:VAP}) and~(\ref{eq:PAV}) note that
$\overline{\epsilon'}\cdot v=0$ and $\epsilon\cdot v'=0$ at zero recoil.

The Isgur-Wise function~$\xi(w)$ is ubiquitous, reappearing, for 
example, in~$h_-(w)$, which is considered in the next section.
Here we are concerned with $v'=v$, and then
\begin{equation}
	\langle c_vJ'|\bar{c}_v i\Gamma^\mu b_v|b_vJ\rangle =
		\omega^\mu \xi(1) = \omega^\mu,
	\label{eq:MJM}
\end{equation}
where $\Gamma^\mu=\gamma^\mu$ or~$\gamma^\mu\gamma_5$ and 
$\omega^\mu=v^\mu$, $v^\mu \overline{\epsilon'}\!\cdot\!\epsilon$, 
$i\overline{\epsilon'}^\mu$, or $-i\epsilon^\mu$, as the case may~be.

\subsubsection{Contributions from $\langle j^{(0)}{\cal L}^{(1)}\rangle$}
\label{app:luke}

The dimension-five interactions in the HQET Lagrangian lead to 
time-ordered products of $j^{(0)}$ with~${\cal O}_2$ and~${\cal O}_B$.
For unequal velocities the matrix elements are parametrized by three 
functions
\begin{eqnarray}
	\int_{-T}^0\!d^4x\, \langle c_{v'}J'| \bar{c}_{v'} \Gamma b_v(0)
	{\cal O}^b_2(x) |b_vJ\rangle^\star & = & 
		-\tr\{\bar{\cal M}_{J'} \Gamma {\cal M}_J\}A_1(w), 
	\label{eq:A1} \\
	\int_{-T}^0\!d^4x\, \langle c_{v'}J'| \bar{c}_{v'} \Gamma b_v(0)
	{\cal O}^b_B(x) |b_vJ\rangle^\star & = & 
		-\tr\{\bar{\cal M}_{J'} \Gamma 
		s^{\alpha\beta}{\cal M}_J A_{\alpha\beta}(v,v')\},
	\label{eq:Aab}
\end{eqnarray}
where, like the chromomagnetic field~$B_{\alpha\beta}$, the 
tensor~$A_{\alpha\beta}(v,v')$ is anti-symmetric and 
$v^\alpha A_{\alpha\beta}(v,v')=0$.
A~general decomposition satisfying these constraints is
\begin{equation}
	A_{\alpha\beta}(v,v') = (\eta i\sigma\eta)_{\alpha\beta} A_3(w)
		+(  i\gamma_{\perp\alpha}v'_{\perp\beta} 
		 - iv'_{\perp\alpha}\gamma_{\perp\beta}) A_2(w).
	\label{eq:A3A2}
\end{equation}
The same functions appear for insertions of~${\cal O}_2^c$ 
and~${\cal O}_B^c$.

One can work out the traces to see how $A_1(w)$ and $A_3(w)$ 
contribute to $h_+(w)$, $h_1(w)$, and $h_{A_1}(w)$.
[$A_2(w)$ contributes to~$h_-(w)$.]
We are, however, mainly interested in the zero-recoil point, $w=1$.
Then the currents become Noether currents, and there are further 
constraints.
With one insertion the starred time-ordered product is identical to 
the connected one:
\begin{eqnarray}
	\langle c_vJ'| \bar{c}_v\Gamma b_v(0)
	{\cal O}_X^b(x)|b_vJ\rangle^\star =
	\langle c_vJ'| \bar{c}_v\Gamma b_v(0) 
	{\cal O}_X^b(x)|b_vJ\rangle_{\text{c}} & = & \nonumber \\
	\langle c_vJ'| \bar{c}_v\Gamma b_v(0) {\cal O}_X^b(x)|b_vJ\rangle 
	 - 
	\langle c_vJ'| \bar{c}_v\Gamma b_v(0)|b_vJ\rangle (&v^0&)^{-1}
	\langle b_vJ|{\cal O}_X^b(x)|b_vJ\rangle ,
	\label{eq:diagonal}
\end{eqnarray}
for $x^0<0$, as in Eq.~(\ref{eq:A1}).
By translation invariance the left-hand side of Eq.~(\ref{eq:A1})
\begin{equation}
	\int\!d^4x\,\langle c_vJ'| T \bar{c}_v\Gamma b_v(y)
	{\cal O}_X^b(x)|b_vJ\rangle^\star = i
	\int\!dx^0\,d^3y\,\langle c_vJ'| T \bar{c}_v\Gamma b_v(y)
	{\cal O}_X^b(x)|b_vJ\rangle^\star .
	\label{eq:charge}
\end{equation}
Taking $\Gamma=iv^0$ and using Eq.~(\ref{eq:MQ=M}) one sees that the
right-hand side of Eq.~(\ref{eq:diagonal}) vanishes identically.
Thus,
\begin{eqnarray}
	\int d^4y\,
	\langle c_vJ'| T\,
	\bar{c}_v \Gamma b_v(0)
	{\cal O}_2(y)
	|b_vJ\rangle^\star & = & 0 , \label{eq:JO2} \\
	\int d^4y\,
	\langle c_vJ'| T\,
	\bar{c}_v \Gamma b_v(0)
	{\cal O}_B(y)
	|b_vJ\rangle^\star & = & 0 , \label{eq:JOB}
\end{eqnarray}
namely $A_1(1)=0$ and $A_3(1)=0$.

These results are properties of heavy-quark symmetry and not of the 
underlying theory.
Usually it is argued that $A_1(1)=A_3(1)=0$ as a consequence of 
current conservation in~QCD.
This line of argument would not have been enough for our purposes, 
because for most choices of~$V^\mu_{\text{lat}}$ current conservation 
fails.
Fortunately, the foregoing argument does not rely on the underlying 
theory; indeed, it is equivalent to the derivation in 
Rayleigh-Schr\"odinger perturbation theory of the Ademollo-Gatto 
theorem.

\subsubsection{Contributions from $\langle j^{(0)}{\cal L}^{(2)}\rangle$}
By the same argument leading to Eqs.~(\ref{eq:JO2}) and~(\ref{eq:JOB})
\begin{eqnarray}
	\int d^4y\,
	\langle c_vJ'| T\,
	\bar{c}_v \Gamma b_v(0)
	{\cal O}_D(y)
	|b_vJ\rangle^\star & = & 0 , \label{eq:JOD} \\
	\int d^4y\,
	\langle c_vJ'| T\,
	\bar{c}_v \Gamma b_v(0)
	{\cal O}_E(y)
	|b_vJ\rangle^\star & = & 0 . \label{eq:JOE}
\end{eqnarray}
The same holds for insertions of the four-quark operators omitted from
Eq.~(\ref{eq:L2}).
Again, this is a property of heavy-quark symmetry and not of the 
underlying theory.

References~\cite{Falk:1993wt,Mannel:1994kv} choose a basis with the 
operator
\begin{equation}
	{\cal Q}_D =
		2\bar{h}_v D_\perp^\mu(-iv\cdot{\cal D})D_{\perp\mu} h_v,
	\label{eq:QD}
\end{equation}
and a similar, spin-dependent operator ${\cal Q}_E$, instead 
of~${\cal O}_D$ and~${\cal O}_E$.
They are related by
\begin{equation}
	 {\cal Q}_D = {\cal O}_D +
	 	\bar{h}_v\loarrow{D}^2_\perp (-iv\cdot{\cal D})h_v +
	 	\bar{h}_v(iv\cdot\loarrow{\cal D})   D^2_\perp h_v ,
	\label{eq:QD=OD}
\end{equation}
up to total derivatives, and similarly for~${\cal Q}_E$.
The additional terms, which superficially vanish by the equations of 
motion, generate contact terms.
Thus,
\begin{eqnarray}
	\int\!d^4y\, \langle c_vJ'| T\, \bar{c}_v \Gamma b_v(0)
		{\cal Q}^b_D(y) |b_vJ\rangle^\star & = & 
	\langle c_vJ'| \bar{c}_v \Gamma D^2_\perp b_v |b_vJ\rangle
		= -\tr\{\bar{\cal M}_{J'} \Gamma {\cal M}_J\} \lambda_1 , 
	\label{eq:JQD} \\
	\int\!d^4y\, \langle c_vJ'| T\, \bar{c}_v \Gamma b_v(0)
		{\cal Q}^b_E(y) |b_vJ\rangle^\star & = & 
	\langle c_vJ'| \bar{c}_v \Gamma {\cal B}  b_v |b_vJ\rangle
		= -\tr\{\bar{\cal M}_{J'} \Gamma 
		s^{\alpha\beta}{\cal M}_Ji\sigma_{\alpha\beta}\} \lambda_2 ,
	 \label{eq:JQE}
\end{eqnarray}
where ${\cal B}=s^{\alpha\beta}B_{\alpha\beta}$,
and $\lambda_1$ and $\lambda_2$ are the same constants (including
any light-sector cutoff effects) as in Eqs.~(\ref{eq:lambda1})
and~(\ref{eq:lambda2}).
The left-hand sides of Eqs.~(\ref{eq:JQD}) and~(\ref{eq:JQE}) were 
parametrized, respectively, with $2B_1(1)$ and $2B_3(1)$ in 
Ref.~\cite{Falk:1993wt} and with $2\Xi_1$ and $2\Xi_3$ in 
Ref.~\cite{Mannel:1994kv}, but the identification with~$\lambda_1$ 
and~$\lambda_2$ was not made.

In the basis employing~${\cal O}_D$ and~${\cal O}_E$, the 
counterpart of these contact terms are the contributions 
$\bar{c}_{v'}\Gamma D_\perp^2 b_v$ and
$\bar{c}_{v'}\Gamma {\cal  B} b_v$ 
to the currents, \emph{cf.}\ Eqs.~(\ref{eq:V02}) and~(\ref{eq:V20}).
In the ${\cal Q}$-basis these currents have coefficients
$(8m^2_{D_\perp^2})^{-1}-(8m^2_D)^{-1}$ and
$(8m^2_{sB})^{-1}-(8m^2_E)^{-1}$.

\subsubsection{Contributions from $\langle j^{(2)}\rangle$}
There are two kinds of of second-order corrections:
those which can be associated with a single leg
and those which involve cross-talk between the legs.
At zero recoil all can be expressed through the parameters~$\lambda_1$ 
and~$\lambda_2$, namely
\begin{equation}
	\langle c_vJ'|\bar{c}_v \Gamma {\cal D}^\alpha 
			{\cal D}^\beta b_v|b_vJ\rangle = 
		-\tr\left\{\bar{\cal M}_{J'} \Gamma {\cal M}_J
		\left[\case{1}{3}\lambda_1\eta^{\alpha\beta}
		  +	  \case{1}{2}\lambda_2 i\sigma^{\alpha\beta}\right]\right\}.
	\label{eq:cDDb}
\end{equation}
By taking $\Gamma$ to be the unit matrix or~$s_{\alpha\beta}$ and 
contracting indices, it is easy to trace back to the 
definitions~(\ref{eq:lambda1}) and~(\ref{eq:lambda2}).
By dimensional analysis, these are the only corrections that can 
arise, even beyond tree level.

The required matrix elements are
\begin{equation}
	\langle c_vJ'|\bar{c}_v i\Gamma_\mu D^2_\perp
		b_v|b_vJ\rangle =
	\langle c_vJ'|\bar{c}_v\loarrow{D}^2_\perp i\Gamma_\mu
		b_v|b_vJ\rangle = \lambda_1 \omega_\mu ,
	\label{eq:MVDDM} 
\end{equation}
\begin{eqnarray}
	\langle c_vJ'|\bar{c}_v i\Gamma_\mu {\cal B}
		b_v|b_vJ\rangle & = & d_{J} \lambda_2 \omega_\mu,
	\label{eq:MASBM}
	\\
	\langle c_vJ'|\bar{c}_v {\cal B} i\Gamma_\mu
		b_v|b_vJ\rangle & = & d_{J'} \lambda_2 \omega_\mu.
	\label{eq:MSBAM}
\end{eqnarray}
At zero recoil
$\langle c_vJ'|\bar{c}_v \loarrow{\cal D}^\alpha%
\Gamma {\cal D}^\beta b_v|b_vJ\rangle =
-\langle c_vJ'|\bar{c}_v \Gamma {\cal D}^\alpha %
{\cal D}^\beta b_v|b_vJ\rangle$,
so~$V^{(1,1)}$ and~$A^{(1,1)}$ have matrix elements
\begin{eqnarray}
	\langle c_vJ|\bar{c}_v
	\loarrow{\kern+0.1em /\kern-0.65em D}_\perp i\gamma_\mu
	{\kern+0.1em /\kern-0.65em D}_\perp
	b_v|b_vJ\rangle & = &
		(\lambda_1 + d_J \lambda_2) \omega_\mu 
	\label{eq:MDVDM} \\
	\langle c_v1|\bar{c}_v
	\loarrow{\kern+0.1em /\kern-0.65em D}_\perp i\gamma_\mu\gamma_5
	{\kern+0.1em /\kern-0.65em D}_\perp
	b_v|b_v0\rangle & = &
	\langle c_v0|\bar{c}_v
	\loarrow{\kern+0.1em /\kern-0.65em D}_\perp i\gamma_\mu\gamma_5
	{\kern+0.1em /\kern-0.65em D}_\perp
	b_v|b_v1\rangle \nonumber \\
		& = & -\case{1}{3} (\lambda_1 + 3 \lambda_2) \omega_\mu
	\label{eq:PDADV} 
\end{eqnarray}
Contributions with $\lambda_1$ ($\lambda_2$) are spin-independent
(spin-dependent).

\subsubsection{Contributions from $\langle{\cal L}^{(1)}j^{(0)}{\cal L}^{(1)}\rangle$}
\label{app:fluke}

Several matrix elements are introduced for double insertions 
of~${\cal L}^{(1)}$.
In the following the short-distance coefficients are stripped off,
leading to insertions of $\int d^4z\,{\cal O}_X^h(z)$, where
$X\in\{2,B\}$ and $h$ labels the heavy flavor.
When the operator comes from the numerator of Eq.~(\ref{eq:ip}) 
the time variable is integrated for $h=b$ over the interval $(-T,0]$
and for $h=c$ over $[0,T)$; when the operator comes from the denominator
the time variable is integrated over the interval~$(-T,T)$. 
After generating all terms the limit $T\to\infty(1-i0^+)$ is taken.

When two interactions occur on the incoming line
\begin{eqnarray}
	\case{1}{2}\int d^4x\,d^4y\,
	\langle c_vJ'| T\,
	\bar{c}_v i\Gamma_\mu b_v(0)
	{\cal O}^b_2(x)
	{\cal O}^b_2(y)
	|b_vJ\rangle^\star & = & \omega_\mu A \label{eq:JO2O2} \\
	\int d^4x\,d^4y\,
	\langle c_vJ'| T\,
	\bar{c}_v i\Gamma_\mu b_v(0)
	{\cal O}^b_2(x)
	{\cal O}^b_B(y)
	|b_vJ\rangle^\star & = & \omega_\mu d_{J} B \label{eq:JO2OB} \\
	\case{1}{2}\int d^4x\,d^4y\,
	\langle c_vJ'| T\,
	\bar{c}_v i\Gamma_\mu b_v(0)
	{\cal O}^b_B(x)
	{\cal O}^b_B(y)
	|b_vJ\rangle^\star & = & \omega_\mu (C_1 + d_{J} C_3) \label{eq:JOBOB}
\end{eqnarray}
and, similarly, when two interactions occur on the outgoing line
\begin{eqnarray}
	\case{1}{2}\int d^4x\,d^4y\,
	\langle c_vJ'| T\,
	{\cal O}^c_2(x)
	{\cal O}^c_2(y)
	\bar{c}_v i\Gamma_\mu b_v(0)
	|b_vJ\rangle^\star & = & \omega_\mu A \label{eq:O2O2J} \\
	\int d^4x\,d^4y\,
	\langle c_vJ'| T\,
	{\cal O}^c_2(x)
	{\cal O}^c_B(y)
	\bar{c}_v i\Gamma_\mu b_v(0)
	|b_vJ\rangle^\star & = & \omega_\mu d_{J'} B \label{eq:O2OBJ} \\
	\case{1}{2}\int d^4x\,d^4y\,
	\langle c_vJ'| T\,
	{\cal O}^c_B(x)
	{\cal O}^c_B(y)
	\bar{c}_v i\Gamma_\mu b_v(0)
	|b_vJ\rangle^\star & = & \omega_\mu (C_1 + d_{J'} C_3) \label{eq:OBOBJ}
\end{eqnarray}
where $\Gamma_\mu=\gamma_\mu$ or $\gamma_\mu\gamma_5$, as the case may
be.
When each line has one interaction
\begin{eqnarray}
	\int d^4x\,d^4y\,
	\langle c_vJ'| T\,
	{\cal O}^c_2(x)
	\bar{c}_v i\Gamma_\mu b_v(0)
	{\cal O}^b_2(y)
	|b_vJ\rangle^\star & = & \omega_\mu D         \label{eq:O2JO2} \\
	\int d^4x\,d^4y\,
	\langle c_vJ'| T\,
	{\cal O}^c_2(x)
	\bar{c}_v i\Gamma_\mu b_v(0)
	{\cal O}^b_B(y)
	|b_vJ\rangle^\star & = & \omega_\mu d_{J} E \label{eq:O2JOB} \\
	\int d^4x\,d^4y\,
	\langle c_vJ'| T\,
	{\cal O}^c_B(x)
	\bar{c}_v i\Gamma_\mu b_v(0)
	{\cal O}^b_2(y)
	|b_vJ\rangle^\star & = & \omega_\mu d_{J'} E \label{eq:OBJO2} 
\end{eqnarray}
again where $\Gamma_\mu=\gamma_\mu$ or $\gamma_\mu\gamma_5$, as the
case may be, and
\begin{eqnarray}
	\int d^4x\,d^4y\,
	\langle c_vJ| T\,
	{\cal O}^c_B(x)
	\bar{c}_v i\gamma_\mu b_v(0)
	{\cal O}^b_B(y)
	|b_vJ\rangle^\star & = & \omega_\mu (R_1 + d_{J} R_2),\label{eq:OBVOB} \\
	\int d^4x\,d^4y\,
	\langle c_v1| T\,
	{\cal O}^c_B(x)
	\bar{c}_v i\gamma_\mu \gamma_5 b(0)
	{\cal O}^b_B(y)
	|b_v0\rangle^\star & = & -\case{1}{3}\omega_\mu (R_1 + 3 R_2).
	\label{eq:OBAOB}
\end{eqnarray}
Here we have used the notation of Ref.~\cite{Mannel:1994kv}.

There are relations between these parameters, which follow solely from 
heavy-quark symmetry and properties of perturbation theory.
Upon expanding Eq.~(\ref{eq:ip}) and sorting terms with like 
coefficients one finds
\begin{eqnarray}
	\int d^4x\,d^4y\,
	\langle c_vJ'|T\, j(0) {\cal O}_X^b(x)
	{\cal O}_Y^b(y) |b_vJ\rangle^\star
	 = &   & \nonumber \\
	\int d^4x\,d^4y\,
	\langle c_vJ'|T\, j(0) {\cal O}_X^b(x)
	{\cal O}_Y^b(y) |b_vJ\rangle_{\text{c}}
	& - & 
	\langle c_vJ'| j(0)|b_vJ\rangle Z_{XY}^\star ,
	\label{eq:MJOOM}
\end{eqnarray}
and
\begin{equation}
	\int d^4x\,d^4y\,
	\langle c_vJ'|{\cal O}_X^c(x) j(0)
	{\cal O}_Y^b(y) |b_vJ\rangle^\star      =
	\int d^4x\,d^4y\,
	\langle c_vJ'|{\cal O}_X^c(x) j(0) 
	{\cal O}_Y^b(y) |b_vJ\rangle_{\text{c}} ,
	\label{eq:MOJOM}
\end{equation}
with limits of integration on the time coordinates as given above.
On the right-hand side of Eq.~(\ref{eq:MJOOM}) the second term is the 
contribution from state renormalization:
\begin{equation}
	Z_{XY}^\star = \frac{1}{v^0}\int_0^T d^4x\,\int_{-T}^0 d^4y\,
	\langle b_vJ|{\cal O}_X^b(x)
	{\cal O}_Y^b(y) |b_vJ\rangle_{\text{c}} ,
	\label{eq:Zstar}
\end{equation}
which is flavor independent.
In Eq.~(\ref{eq:MJOOM}) the operator~$j$ is left-most for all time 
orderings.
When $j$ is a Noether charge one can apply Eq.~(\ref{eq:charge})
to show that the connected term vanishes, leaving only the term from
state renormalization.
In Eq.~(\ref{eq:MOJOM}) the operator~$j$ is in the middle for all time 
orderings.
When~$j$ is a Noether charge, however, the right-hand side can be 
reduced to the same quantity as in the state renormalization.
Inserting complete sets of states on both sides of $j$, and noting that 
Eq.~(\ref{eq:charge}) applies equally well to excited states, one finds
\begin{equation}
	\int d^4x\,d^4y\,
	\langle c_vJ'|{\cal O}_X^c(x) j(0)
	{\cal O}_Y^b(y) |b_vJ\rangle^\star =
	\langle c_vJ'| j(0)|b_vJ\rangle Z_{XY}^\star ,
	\label{eq:MOJOM=MJOOM}
\end{equation}
making use of the flavor independence of~$Z_{XY}^\star$.
Apart from a sign, therefore, the two kinds of $\star$-products are 
the same, and
\begin{eqnarray}
	 A  = -\case{1}{2}  D,  \label{eq:A=D}   \\
	 B  =       -       E,  \label{eq:B=E}   \\
	C_1 = -\case{1}{2} R_1, \label{eq:C1=R1} \\
	C_3 = -\case{1}{2} R_2. \label{eq:C3=R2}
\end{eqnarray}
These identities leave only four parameters.
To my knowledge they have not been derived before.
Since they do not depend on the underlying theory, they hold also for 
continuum~QCD.

\subsection{Near zero recoil: $h_-(1)$}
\label{app:near0}

At zero recoil several matrix elements vanish, but they are precisely
of the type leading to~$h_-(w)$ in Eq.~(\ref{eq:h+h-}).
To extract~$h_-(1)$ one must take $v'\neq v$ while evaluating
matrix elements, read off the form factor, and then set~$w$ to~1.
This subsection works out the relevant matrix elements, those of the
dimension-four currents, the dimension-five currents 
$\bar{c}_{v'}i\gamma_\mu i{\kern+0.1em /\kern-0.65em E} b_v$ and 
$\bar{c}_{v'}i{\kern+0.1em /\kern-0.65em E}'i\gamma_\mu b_v$, and 
time-ordered products of dimension-four currents with~${\cal L}^{(1)}$.

\subsubsection{Contributions from $\langle j^{(1)}\rangle$}

For the matrix elements
$\langle c_{v'}0|\bar{c}_{v'}i\gamma_\mu%
{\kern+0.1em /\kern-0.65em D}_\perp b_v|b_v0\rangle$ and
$\langle c_{v'}0|\bar{c}_{v'}%
\loarrow{\kern+0.1em /\kern-0.65em D}_{\perp'}%
i\gamma_\mu b_v|b_v0\rangle$
one starts with the matrix element
\begin{equation}
	\langle c_{v'}J'|\bar{c}_{v'}\Gamma
	{\cal D}^\alpha b_v|b_vJ\rangle = 
		-\tr\{\bar{\cal M}_{J'} \Gamma {\cal M}_J i\xi^\alpha(v,v')\}
	\label{eq:cGDb}
\end{equation}
where~$\xi^\alpha$ parametrizes the light degrees of freedom.
The equation of motion $(-iv\cdot{\cal D})b_v=0$ implies that
$v_\alpha\xi^\alpha(v,v')=0$, leaving two independent form factors
\begin{equation}
	\xi^\alpha(v,v') =
		v_\perp^{\prime\alpha}\xi_2(w) - i\gamma_\perp^\alpha\xi_3(w).
	\label{eq:3xi}
\end{equation}
A further constraint on $\xi^\alpha(v,v')$ comes from the 
``integration-by-parts'' identity
\begin{equation}
	\langle c_{v'}J'|\bar{c}_{v'}\loarrow{\cal D}^\alpha
		\Gamma b_v|b_vJ\rangle + 
	\langle c_{v'}J'|\bar{c}_{v'}\Gamma {\cal D}^\alpha b_v|b_vJ\rangle
		= -i\bar{\Lambda}(v'-v)^\alpha
			\langle c_{v'}J'|\bar{c}_{v'}\Gamma b_v|b_vJ\rangle ,
	\label{eq:parts}
\end{equation}
where ${\cal D}b_v=(D-im_{1b}v)b_v$ and
$\bar{c}_{v'}\loarrow{\cal D}=\bar{c}_{v'}(\loarrow{D}+im_{1c}v')$.
The first matrix element
\begin{equation}
	\langle c_{v'}J'|\bar{c}_{v'}\loarrow{\cal D}^\alpha 
		\Gamma b_v|b_vJ\rangle = 
		-\tr\{\bar{\cal M}_{J'} \Gamma {\cal M}_J
			[-i\overline{\xi^\alpha(v',v)}]\},
	\label{eq:cDGb}
\end{equation}
where $\overline{\xi^\alpha(v',v)}=%
\gamma^4[\xi^\alpha(v',v)]^\dagger\gamma^4$.
Substituting traces for matrix elements in Eq.~(\ref{eq:parts}) yields
the relation
\begin{equation}
	(w+1)\xi_2(w) + \xi_3(w) = - \bar{\Lambda}\xi(w),
	\label{eq:xi2}
\end{equation}
which can be used to eliminate~$\xi_2(w)$.
In Eq.~(\ref{eq:xi2}) the constant $\bar{\Lambda}$ and the function
$\xi(w)$ are the same---including lattice artifacts of the light
degrees of freedom---as in Eqs.~(\ref{eq:Lambda}) and~(\ref{eq:xi}),
respectively.
Evaluating the traces of interest and using Eq.~(\ref{eq:xi2}) one 
finds
\begin{eqnarray}
	\langle c_{v'}0|\bar{c}_{v'}i\gamma_\mu 
		{\kern+0.1em /\kern-0.65em D}_\perp b_v|b_v0\rangle 
	& = & \langle c_{v'}0|\bar{c}_{v'}
		\loarrow{\kern+0.1em /\kern-0.65em D}_{\perp'}
		i\gamma_\mu  b_v|b_v0\rangle
	\nonumber \\
	& = & - \case{1}{2}(v'-v)_\mu [2\xi_3(w) - \bar{\Lambda}\xi(w)],
	\label{eq:PV3P}
\end{eqnarray}
There is no contribution to~$h_+(w)$, and the vector-to-vector matrix
elements make no contribution to~$h_1(w)$, just to other form factors
that are not considered in this paper.
An equivalent analysis appears in Ref.~\cite{Falk:1993wt}.
The only significant addition is to extend to lattice QCD the
identification of~$\bar{\Lambda}\xi(w)$ in Eq.~(\ref{eq:xi2})
with the quantities in Eqs.~(\ref{eq:Lambda}) and~(\ref{eq:xi}).

\subsubsection{Contributions from $\langle j^{(2)}\rangle$}
To obtain all of the second-order corrections to the current one can
start with
\begin{equation}
	\langle c_{v'}J'|\bar{c}_{v'}\loarrow{\cal D}^\alpha
		\Gamma{\cal D}^\beta b_v|b_vJ\rangle =
	-\tr\{\bar{\cal M}_{J'} \Gamma {\cal M}_J
	[-\lambda^{\alpha\beta}(v,v')]\}.
	\label{eq:cDGDb}
\end{equation}
The equations of motion $(-iv\cdot{\cal D})b_v=0$ and
$\bar{c}_{v'}(iv'\cdot\loarrow{\cal D})=0$ imply that
$\lambda^{\alpha\beta}(v,v')v_\beta=0$ and
$v'_\alpha\lambda^{\alpha\beta}(v,v')=0$, and symmetry
under exchanging final and initial states implies that
$\overline{\lambda^{\beta\alpha}(v',v)}=\lambda^{\alpha\beta}(v,v')$,
leaving four independent form factors,
\begin{equation}
	\lambda^{\alpha\beta}(v,v') =
	{\eta'}^\alpha_\gamma [\case{1}{3} g^{\gamma\delta} \lambda_1(w) +
	\case{1}{2}i\sigma^{\gamma\delta} \lambda_2(w)] \eta_\delta^\beta +
	v_{\perp'}^\alpha {v'}_\perp^\beta \lambda_3(w) +
	[i\gamma_{\perp'}^\alpha {v'}_\perp^\beta +
 	  v_{\perp'}^\alpha   i\gamma_\perp^\beta] \lambda_4(w).
	\label{eq:lambda}
\end{equation}
The pre-factors for the first two form factors are chosen so that
$\lambda_1(1)=\lambda_1$ and $\lambda_2(1)=\lambda_2$ are the constants
in Eq.~(\ref{eq:cDDb}).

The matrix elements needed for~$h_-(w)$ are
$\langle c_{v'}0|%
\bar{c}_{v'}i\gamma_\mu i{\kern+0.1em /\kern-0.65em E} b_v%
|b_v0\rangle$ and
$\langle c_{v'}0|%
\bar{c}_{v'}i{\kern+0.1em /\kern-0.65em E}'i\gamma_\mu b_v%
|b_v0\rangle$.
They are related to Eq.~(\ref{eq:cDGDb}) by the identity
\begin{equation}
	\langle c_{v'}J'|\bar{c}_{v'}\Gamma
	{\cal D}^\alpha {\cal D}^\beta b_v|b_vJ\rangle =
	 - i\bar{\Lambda}(v'-v)^\alpha
		\langle c_{v'}J'|\bar{c}_{v'}\Gamma
		{\cal D}^\beta b_v|b_vJ\rangle
	 - \langle c_{v'}J'|\bar{c}_{v'}\loarrow{\cal D}^\alpha
		\Gamma {\cal D}^\beta b_v|b_vJ\rangle
	\label{eq:parts2} 
\end{equation}
and the definitions 
$i{\kern+0.1em /\kern-0.65em E} =%
-iv_\alpha [{\cal D}^\alpha,{\cal D}_\beta]\gamma^\beta_\perp$,
$i{\kern+0.1em /\kern-0.65em E}'=%
-iv'_\alpha[{\cal D}^\alpha,{\cal D}_\beta]\gamma^\beta_{\perp'}$.
Evaluating the traces one finds
\begin{eqnarray}
	\langle c_{v'}0|\bar{c}_{v'}i\gamma^\mu
	i{\kern+0.1em /\kern-0.65em E} b_v|b_v0\rangle 
	& = &
	\langle c_{v'}0|\bar{c}_{v'}i{\kern+0.1em /\kern-0.65em E}' 
	i\gamma^\mu b_v|b_v0\rangle \nonumber \\ & = &
	-\case{1}{2}(v'-v)^\mu \left\{\case{1}{2}(w+1)\lambda(w)
		+ (w-1)\bar{\Lambda}\left[2\xi_3(w)-\bar{\Lambda}\xi(w)\right]
	\right\}
\end{eqnarray}
where
\begin{equation}
	\lambda(w) = \case{2}{3} w \lambda_1(w) + (3-w) \lambda_2(w) 
		- 2 (w^2-1) \lambda_3(w) + 8 (w-1) \lambda_4(w).
\end{equation}
At $w=1$, $\lambda(1)=\case{2}{3}(\lambda_1+3\lambda_2)$.

\subsubsection{Contributions from $\langle j^{(1)}{\cal L}^{(1)}\rangle$}

The time-ordered products of interest are
$\int d^4y\, \langle c_{v'}J'|%
T\,\bar{c}_{v'}\Gamma {\kern+0.1em /\kern-0.65em D}_\perp b_v(x)%
{\cal O}_X^f(y) |b_vJ\rangle^\star$ and
$\int d^4y\, \langle c_{v'}J'|%
T\,\bar{c}_{v'}\loarrow{\kern+0.1em /\kern-0.65em D}_{\perp'}\Gamma b_v(x)%
{\cal O}_X^f(y) |b_vJ\rangle^\star$,
where $X\in\{2,B\}$ and $f\in\{c,b\}$.
As before it is helpful to consider matrix elements with
${\kern+0.1em /\kern-0.65em D}_\perp$ replaced with ${\cal D}$
and derive constraints from the equations of motion and from
``integrating by parts.''
This is a bit trickier now, with derivatives acting under the 
time-ordered product.

The equations of motion imply the identities
\begin{eqnarray}
	\int d^4y\,
	\langle c_{v'}J'|
		T\,\bar{c}_{v'}\Gamma (-iv\cdot{\cal D})b_v(x) {\cal O}_2^b(y)
	|b_vJ\rangle^\star & = &
	\langle c_{v'}J'|
		\bar{c}_{v'}\Gamma D^2_\perp b_v(x)
	|b_vJ\rangle , \label{eq:eom2b} \\
	\int d^4y\,
	\langle c_{v'}J'|
		T\,\bar{c}_{v'}\Gamma (-iv\cdot{\cal D})b_v(x) {\cal O}_B^b(y)
	|b_vJ\rangle^\star & = &
	\langle c_{v'}J'|
		\bar{c}_{v'}\Gamma {\cal B} b_v(x)
	|b_vJ\rangle , \label{eq:eomBb} \\
	\int d^4y\,
	\langle c_{v'}J'| T\,{\cal O}_2^c(y)
		\bar{c}_{v'}(iv'\cdot\loarrow{\cal D})\Gamma b_v(x)
	|b_vJ\rangle^\star & = &
	\langle c_{v'}J'|
		\bar{c}_{v'}\loarrow{D}^2_{\perp'} \Gamma b_v(x)
	|b_vJ\rangle , \label{eq:eom2c} \\
	\int d^4y\,
	\langle c_{v'}J'| T\,{\cal O}_B^c(y)
		\bar{c}_{v'}(iv'\cdot\loarrow{\cal D})\Gamma b_v(x)
	|b_vJ\rangle^\star & = &
	\langle c_{v'}J'|
		\bar{c}_{v'}{\cal B}'\Gamma b_v(x)
	|b_vJ\rangle . \label{eq:eomBc} 
\end{eqnarray}
The contact terms on the right-hand side were omitted from Eqs.~(4.27)
of Ref.~\cite{Falk:1993wt} but do appear, for example, in Eq.~(A21)
of Ref.~\cite{Falk:1994dh}.
They arise from a careful definition of the $T$-product for operators
containing time derivatives.
A~helpful mnemonic for checking them is to note that
\begin{eqnarray}
	(-iv\cdot{\cal D})T\,b_v(x) b_v(y) & = & 
		\delta^{(4)}(x-y), \label{eq:eompropb} \\
	T\,c_{v'}(y)\bar{c}_{v'}(x)(iv'\cdot\loarrow{\cal D}) & = &
		\delta^{(4)}(y-x). \label{eq:eompropc} 
\end{eqnarray}
Further identities come from taking the derivative
$\partial^\rho\langle D_{v'}|\bar{c}_{v'}\Gamma b_v|B_v\rangle$
between fully dressed states, and generating the expansion.
This leads to
\begin{eqnarray}
	\int d^4y\,
	\langle c_{v'}J'|
		T\,[\bar{c}_{v'}\Gamma {\cal D}^\rho b_v(x) +
		\bar{c}_{v'}\loarrow{\cal D}^\rho \Gamma b_v(x)]{\cal O}_2^b(y) 
	|b_vJ\rangle^\star & = &
	\label{eq:ibp2b} \\
	-i\bar{\Lambda}(v'-v)^\rho
	\int d^4y\,
	\langle c_{v'}J'|
		T\,\bar{c}_{v'}\Gamma b_v(x) {\cal O}_2^b(y) 
	|b_vJ\rangle^\star & - & 
	i\lambda_1 v^\rho \langle c_{v'}J'|
		\bar{c}_{v'}\Gamma b_v(x)
	|b_vJ\rangle ,
	\nonumber \\ 
	\int d^4y\,
	\langle c_{v'}J'|
		T\,[\bar{c}_{v'}\Gamma {\cal D}^\rho b_v(x) +
		\bar{c}_{v'}\loarrow{\cal D}^\rho \Gamma b_v(x)]{\cal O}_B^b(y) 
	|b_vJ\rangle^\star & = &
	\label{eq:ibpBb} \\
	-i\bar{\Lambda}(v'-v)^\rho
	\int d^4y\,
	\langle c_{v'}J'|
		T\,\bar{c}_{v'}\Gamma b_v(x) {\cal O}_B^b(y) 
	|b_vJ\rangle^\star & - & 
	id_{J}\lambda_2 v^\rho \langle c_{v'}J'|
		\bar{c}_{v'}\Gamma b_v(x)
	|b_vJ\rangle ,
	\nonumber \\ 
	\int d^4y\,
	\langle c_{v'}J'|
		T\,{\cal O}_2^c(y)[\bar{c}_{v'}\Gamma {\cal D}^\rho b_v(x) +
		\bar{c}_{v'}\loarrow{\cal D}^\rho \Gamma b_v(x)] 
	|b_vJ\rangle^\star & = &
	\label{eq:ibp2c} \\
	-i\bar{\Lambda}(v'-v)^\rho
	\int d^4y\,
	\langle c_{v'}J'|
		T\,{\cal O}_2^c(y) \bar{c}_{v'}\Gamma b_v(x) 
	|b_vJ\rangle^\star & + & 
	i\lambda_1 v'^\rho \langle c_{v'}J'|
		\bar{c}_{v'}\Gamma b_v(x)
	|b_vJ\rangle ,
	\nonumber \\ 
	\int d^4y\,
	\langle c_{v'}J'|
		T\,{\cal O}_B^c(y)[\bar{c}_{v'}\Gamma {\cal D}^\rho b_v(x) +
		\bar{c}_{v'}\loarrow{\cal D}^\rho \Gamma b_v(x)] 
	|b_vJ\rangle^\star & = &
	\label{eq:ibpBc} \\
	-i\bar{\Lambda}(v'-v)^\rho
	\int d^4y\,
	\langle c_{v'}J'|
		T\,{\cal O}_B^c(y) \bar{c}_{v'}\Gamma b_v(x) 
	|b_vJ\rangle^\star & + & 
	i d_{J'}\lambda_2 v'^\rho \langle c_{v'}J'|
		\bar{c}_{v'}\Gamma b_v(x)
	|b_vJ\rangle .
	\nonumber
\end{eqnarray}
These identities do not agree with analogous ones from combining
Eqs.~(C4) and~(C5) of Ref.~\cite{Falk:1993wt}.
Remarkably, Eq.~(C5) of Ref.~\cite{Falk:1993wt} contains the contact
terms omitted from Eq.~(4.27) of Ref.~\cite{Falk:1993wt}.

Once again the time-ordered products are parametrized by form factors.
Consider first the case with the kinetic operator.
It is enough to present the details for~${\cal O}^b_2$.
One may write
\begin{eqnarray}
	\int d^4y\,\langle c_{v'}J'|T\,\bar{c}_{v'}\Gamma
	{\cal D}^\alpha b_v {\cal O}^b_2(y)|b_vJ\rangle^\star & = & 
		-\tr\{\bar{\cal M}_{J'} \Gamma {\cal M}_J i\Xi^\alpha(v,v')\},
	\label{eq:cGDbO2} \\
	\int d^4y\,\langle c_{v'}J'|T\,\bar{c}_{v'}\loarrow{\cal D}^\alpha 
		\Gamma b_v {\cal O}^b_2(y)|b_vJ\rangle^\star & = & 
		-\tr\{\bar{\cal M}_{J'} \Gamma {\cal M}_J[-iF^\alpha(v,v')]\}.
	\label{eq:cDGbO2}
\end{eqnarray}
In the first case, the equation of motion~(\ref{eq:eom2b}) implies
$v\cdot\Xi=\phi_0$, where
\begin{equation}
	\phi_0(w) = \case{1}{3}(2+w^2)\lambda_1(w) 
		-(w-1)\{ w[\case{1}{2}\lambda_2(w) 
			+ (w+1)\lambda_3(w) - 2\lambda_4(w)]
		+ \bar{\Lambda}^2\xi(w) \}
	\label{eq:phi0(w)}
\end{equation}
is obtained from Eqs.~(\ref{eq:lambda}) and~(\ref{eq:parts2}).
Thus, $\Xi^\alpha$ has a decomposition
\begin{equation}
	\Xi^\alpha(v,v') = - v^\alpha\phi_0(w) + 
		v_\perp^{\prime\alpha}\Xi_2(w) - i\gamma^\alpha_\perp \Xi_3(w)
	\label{eq:Xi}
\end{equation}
similar to~$\xi^\alpha$ but with $-\phi_0(w)$ multiplying $v^\alpha$.
On the other hand, the equation of motion still implies
$v'\cdot F=0$, so $F^\alpha$ has the decomposition
\begin{equation}
	F^\alpha(v,v') = 
		v_{\perp'}^\alpha  F_1(w) - i\gamma^\alpha_{\perp'}F_3(w).
	\label{eq:F}
\end{equation}
similar to $\overline{\xi^\alpha(v',v)}$.
The form factors~$\Xi_2$, $F_1$, and $F_3$ can be eliminated, because
the identity~(\ref{eq:ibp2b}) implies
\begin{eqnarray}
	(w+1)\Xi_2 + \Xi_3 & = & -w\tilde{\phi}_0 - \bar{\Lambda}A_1 ,
	\label{eq:Xi2}  \\
	(w+1)  F_1 +   F_3 & = &   \tilde{\phi}_0 - \bar{\Lambda}A_1 ,
	\label{eq:F2}  \\
	               F_3 & = & \Xi_3 ,
	\label{eq:F3}
\end{eqnarray}
where
\begin{equation}
	\tilde{\phi}_0(w) =\frac{\phi_0(w) - \lambda_1\xi(w)}{w-1}.
	\label{eq:tildephi0(w)}
\end{equation}
At zero recoil the new constants that can arise are~$\Xi_3(1)$ and, 
denoting differentiation with respect to~$w$ by a dot, 
$\tilde{\phi}_0(1)=\dot{\phi}_0(1)-\lambda_1\dot{\xi}(1)$.
(Recall that $A_1(1)=0$, as a consequence of heavy-quark flavor
symmetry.)

Evaluating the traces for~$h_-(w)$, one finds
\begin{eqnarray}
	\int d^4y\, \langle c_{v'}0| T\,\bar{c}_{v'}i\gamma^\mu 
		{\kern+0.1em /\kern-0.65em D}_\perp b_v(x) {\cal O}_2^b(y) 
	|b_v0\rangle^\star & = & 
	\int d^4y\, \langle c_{v'}0| T\,\bar{c}_{v'} {\cal O}_2^c(y) 
		\loarrow{\kern+0.1em /\kern-0.65em D}_{\perp'} i\gamma^\mu b_v(x)
	|b_v0\rangle^\star \nonumber \\
		& = & - \case{1}{2}(v'-v)^\mu 
		[2\Xi_3(w) - \bar{\Lambda}A_1(w) - w\tilde{\phi}_0(w)] ,
	\label{eq:TDO2b} \\
	\int d^4y\, \langle c_{v'}0| T\,\bar{c}_{v'}
		\loarrow{\kern+0.1em /\kern-0.65em D}_{\perp'} i\gamma^\mu b_v(x)
	{\cal O}_2^b(y) |b_v0\rangle^\star & = &
	\int d^4y\, \langle c_{v'}0| T\,\bar{c}_{v'}{\cal O}_2^c(y) i\gamma^\mu 
		{\kern+0.1em /\kern-0.65em D}_\perp b_v(x) 
	|b_v0\rangle^\star \nonumber \\ 
		& = & - \case{1}{2}(v'-v)^\mu
		[2\Xi_3(w) - \bar{\Lambda}A_1(w) +  \tilde{\phi}_0(w)] .
	\label{eq:TDO2c}
\end{eqnarray}
Matrix elements of this kind make no contribution to~$h_+(w)$,
$h_1(w)$, or~$h_{A_1}(w)$.

Finally there are the time-ordered products with the chromomagnetic
energy.
It is enough to show the details for ${\cal O}_B^b$.
When the derivative acts on~$b_{v}$,
\begin{equation}
	\int d^4y\, \langle c_{v'}J'|T\,\bar{c}_{v'}\Gamma
	{\cal D}^\rho b_v(x) {\cal O}^b_B(y)|b_vJ\rangle^\star = 
		-\tr\{\bar{\cal M}_{J'} \Gamma s^{\alpha\beta}
		{\cal M}_J i{\Xi^\rho}_{\alpha\beta}(v,v')\},
	\label{eq:cGDbOB}
\end{equation}
and when the derivative acts on~$\bar{c}_{v'}$,
\begin{equation}
	\int d^4y\, \langle c_{v'}J'|T\,\bar{c}_{v'}\loarrow{\cal D}^\rho 
	\Gamma b_v (x) {\cal O}^b_B(y)|b_vJ\rangle^\star = 
		-\tr\{\bar{\cal M}_{J'} \Gamma s^{\alpha\beta}
		{\cal M}_J [-i{F^\rho}_{\alpha\beta}(v,v')]\}.
	\label{eq:cDGbOB}
\end{equation}
The tensors ${\Xi^\rho}_{\alpha\beta}$ and ${F^\rho}_{\alpha\beta}$
inherit properties from the chromomagnetic field $B_{\alpha\beta}$:
they are antisymmetric on the lower indices, and
${\Xi^\rho}_{\alpha\beta}v^\beta={F^\rho}_{\alpha\beta}v^\beta=0$.
From the equations of motion $v'_\rho{F^\rho}_{\alpha\beta}=0$ and
$v_\rho{\Xi^\rho}_{\alpha\beta}=\phi_{\alpha\beta}$, where
\begin{equation}
	\phi_{\alpha\beta}(v,v') =
		\eta_{\alpha\mu}[\lambda^{\mu\nu}(v,v') - 
		\lambda^{\nu\mu}(v,v')] \eta_{\nu\beta} +
		\bar{\Lambda}[v'_{\perp\alpha}\xi_\beta(v,v') -
			v'_{\perp\beta}\xi_\alpha(v,v')].
	\label{eq:phiab}
\end{equation}
Substituting Eq.~(\ref{eq:lambda}) into Eq.~(\ref{eq:phiab})
\begin{equation}
	\phi_{\alpha\beta}(v,v') = 
		(\eta i\sigma\eta)_{\alpha\beta} \phi_3(w) -
		(  i\gamma_{\perp\alpha}v'_{\perp\beta} 
		 - iv'_{\perp\alpha}\gamma_{\perp\beta}) \phi_2(w),
\end{equation}
where $\phi_3(w)=\lambda_2(w)$ and
$\phi_2(w)=%
-\case{1}{2}\lambda_2(w)-(w+1)\lambda_4(w)-\bar{\Lambda}\xi_3(w)$.
The constraints on~${\Xi^\rho}_{\alpha\beta}$
and~${F^\rho}_{\alpha\beta}$ lead to the decompositions
\begin{eqnarray}
	{\Xi^\rho}_{\alpha\beta}(v,v') & = & 
	(\eta i\sigma\eta)_{\alpha\beta}\left[
		- v^\rho\phi_3
		+ v_\perp^{\prime\rho}\Xi_8
		- i\gamma_\perp^\rho  \Xi_9\right] \nonumber \\
	& - & (i\gamma_{\perp\alpha}v'_{\perp\beta} 
		 - v'_{\perp\alpha}i\gamma_{\perp\beta})\left[
		- v^\rho\phi_2
		+ v_\perp^{\prime\rho}\Xi_5
		+ i\gamma_\perp^\rho  \Xi_6\right] \label{eq:Xirab} \\
	& + & (\eta^\rho_\alpha v'_{\perp\beta}
		  -\eta^\rho_\beta v'_{\perp\alpha})\Xi_{10}
		+ (\eta^\rho_\alpha i\gamma_{\perp\beta}
		  -\eta^\rho_\beta i\gamma_{\perp\alpha})\Xi_{11}, \nonumber
\end{eqnarray}
and
\begin{eqnarray}
	{F^\rho}_{\alpha\beta}(v,v') & = & 
	(\eta i\sigma\eta)_{\alpha\beta}\left[
		  v_{\perp'}^\rho F_7
		- i\gamma_{\perp'}^\rho  F_9\right] \nonumber \\
	& - & (i\gamma_{\perp\alpha}v'_{\perp\beta} 
		 - v'_{\perp\alpha}i\gamma_{\perp\beta})\left[
		  v_{\perp'}^\rho F_4
		+ i\gamma_{\perp'}^\rho  F_6\right] \label{eq:Frab} \\
	& + & \eta^{\prime\rho}_\sigma(\eta^\sigma_\alpha v'_{\perp\beta}
		  -\eta^\sigma_\beta v'_{\perp\alpha})F_{10}
		+ \eta^{\prime\rho}_\sigma(\eta^\sigma_\alpha i\gamma_{\perp\beta}
		  -\eta^\sigma_\beta i\gamma_{\perp\alpha})F_{11}. \nonumber
\end{eqnarray}
The subscripts are chosen as in Ref.~\cite{Falk:1993wt}.

The identity~(\ref{eq:ibpBb}) can be applied to eliminate~$\Xi_5$, 
$\Xi_8$, and all~$F$s:
\begin{eqnarray}
	F_k = \Xi_k , &   & k\in\{6,9,10,11\}, \label{eq:FX} \\
	(w^2-1)\Xi_5 & = & -w\phi_2 - (w+1)\Xi_6 -  \Xi_{11}  
				+ (w-1)\bar{\Lambda}A_2 , \label{eq:Xi5} \\
	(w^2-1)  F_4 & = &   \phi_2 + (w+1)\Xi_6 + w\Xi_{11}   
				+ (w-1)\bar{\Lambda}A_2 , \label{eq:F4}  \\
	(w+1)\Xi_8 & = & -w\tilde{\phi}_3 - \Xi_9 - \bar{\Lambda}A_3 ,
	\label{eq:Xi8} \\
	(w+1)  F_7 & = &   \tilde{\phi}_3 - \Xi_9 - \bar{\Lambda}A_3 ,
	\label{eq:F7}
\end{eqnarray}
where
\begin{equation}
	\tilde{\phi}_3(w) = \frac{\phi_3(w) - \lambda_2\xi(w)}{w-1}.
	\label{eq:tildephi3(w)}
\end{equation}
Each of Eqs.~(\ref{eq:Xi5}) and~(\ref{eq:F4}) implies 
$2\Xi_6(1)+\Xi_{11}(1)=-\phi_2(1)$.

Evaluating the traces for~$h_-(w)$, one finds
\begin{eqnarray}
	\int d^4y\, \langle c_{v'}0| T\,\bar{c}_{v'}i\gamma^\mu 
		{\kern+0.1em /\kern-0.65em D}_\perp b_v(x) {\cal O}_B^b(y) 
	|b_v0\rangle^\star & = & 
	\int d^4y\, \langle c_{v'}0| T\,\bar{c}_{v'} {\cal O}_B^c(y) 
		\loarrow{\kern+0.1em /\kern-0.65em D}_{\perp'} i\gamma^\mu b_v(x)
	|b_v0\rangle^\star \nonumber \\
		& = & - \case{1}{2}(v'-v)^\mu [2\Xi_-(w) - w\tilde{\phi}_-(w)] ,
	\label{eq:b3bB} \\
	\int d^4y\, \langle c_{v'}0| T\,\bar{c}_{v'}
		\loarrow{\kern+0.1em /\kern-0.65em D}_{\perp'} i\gamma^\mu b_v(x)
	{\cal O}_B^b(y) |b_v0\rangle^\star & = &
	\int d^4y\, \langle c_{v'}0| T\,\bar{c}_{v'}{\cal O}_B^c(y) i\gamma^\mu 
		{\kern+0.1em /\kern-0.65em D}_\perp b_v(x) 
	|b_v0\rangle^\star \nonumber \\ 
		& = & - \case{1}{2}(v'-v)^\mu [2\Xi_-(w) +  \tilde{\phi}_-(w)] ,
	\label{eq:c3bB}
\end{eqnarray}
where
\begin{eqnarray}
	\Xi_-   & = & 3\Xi_9 + (w+1)(2\Xi_6+\Xi_{10}) - 2\Xi_{11}
		- \case{3}{2}\bar{\Lambda}A_3 - (w-1)\bar{\Lambda}A_2 .
	\label{eq:Xi-} \\
	\tilde{\phi}_- & = & 3\tilde{\phi}_3 - 2\phi_2
	\label{eq:tildephi-}
\end{eqnarray}
At zero recoil the new constants that can arise are~$\Xi_6(1)$, 
$\Xi_9(1)$, $\Xi_{10}(1)$, and $\tilde{\phi}_3(1)$.
(Note that $A_2(1)$ drops out, and recall that $A_3(1)=0$ as a
consequence of heavy-quark flavor symmetry.)
As in Eqs.~(\ref{eq:TDO2b}) and~(\ref{eq:TDO2c}), matrix elements of
this kind make no contribution to~$h_+(w)$, $h_1(w)$, or~$h_{A_1}(w)$.
In~$h_-(1)$ they reduce to two constants
\begin{eqnarray}
	\Xi_-(1)   & = & 3\Xi_9(1) + 8 \Xi_6(1) + 2\Xi_{10}(1) + 2\phi_2(1) ,
	\label{eq:Xi-(1)} \\
	\tilde{\phi}_-(1) & = & 3[\dot{\phi}_3(1) - \lambda_2\dot{\xi}(1)]
	 - 2\phi_2(1),
	\label{eq:tildephi-(1)}
\end{eqnarray}
which are needed in Eq.~(\ref{eq:X100}).

\end{document}